\documentclass[onecolumn]{IEEEtran}

\usepackage{subfigure}
\usepackage{amssymb,amsmath}
\usepackage{textcomp}
\usepackage{hyperref}
\usepackage{graphicx}
\usepackage[justification=centering]{caption}
\usepackage{float}
\usepackage{tablefootnote}
\usepackage{setspace}
\usepackage{array}
\usepackage[para,online,flushleft]{threeparttable}

		\usepackage{tikz} 

\usetikzlibrary{decorations.pathreplacing}
\usetikzlibrary{positioning}
\usetikzlibrary{calc}
\usetikzlibrary{fit}
\usetikzlibrary{matrix}
\usetikzlibrary{external}

\DeclareMathOperator*{\argmin}{arg\,min}
\newcommand{\Mod}[1]{\ (\text{mod}\ #1)}

\def\beginofproof{\indent {\it Proof: }}
\def\endofproof{\hfill\rule{6pt}{6pt}}
\newtheorem{theorem}{Theorem}
\newtheorem{lemma}{Lemma}

\newtheorem{corollary}{Corollary}
\newtheorem{remark}{Remark}

\ifCLASSINFOpdf
\else
\fi

\begin{document}
%
\title{Secure Transmission on the Two-hop Relay Channel with Scaled Compute-and-Forward}
%
%
%

\author{
\IEEEauthorblockN{
Zhijie~Ren,
Jasper~Goseling,
Jos~H.~Weber,
and Michael~Gastpar%
\thanks{Z. Ren is with the Department of Intelligent Systems at Delft University of Technology, 2628CD, Delft, the Netherlands (e-mail: z.ren@tudelft.nl).}%
\thanks{J. Goseling is with the Department of Applied Mathematics at University of Twente, 7522 NH, Enschede, the Netherlands (e-mail: j.goseling@utwente.nl).}%
\thanks{J.H. Weber is with the Department of Intelligent Systems at Delft University of Technology, 2628CD, Delft, the Netherlands (e-mail: j.h.weber@tudelft.nl).}%
\thanks{M. Gastpar is with the Laboratory of Information and Networked Systems at Ecole polytechnique f\'ed\'erale de Lausanne, CH-1015, Lausanne and also with the Department of Intelligent Systems at Delft University of Technology, 2628CD, Delft, the Netherlands (e-mail: michael.gastpar@epfl.ch).}%
\thanks{The work in this paper was partially presented at the Information Theory Workshop, Jerusalem, Israel, Apr., 2015.}%
}

}
\maketitle

\begin{abstract}
In this paper, we consider communication on a two-hop channel, in which a source wants to send information reliably and securely to the destination via a relay. We consider both the untrusted relay case and the external eavesdropper case. In the untrusted relay case, the relay behaves as an eavesdropper and there is a cooperative node which sends a jamming signal to confuse the relay when the it is receiving from the source. We propose two secure transmission schemes using the scaled compute-and-forward technique. One of the schemes is based on a random binning code and the other one is based on a lattice chain code. It is proved that in either the high Signal-to-Noise-Ratio (SNR) scenario and/or the restricted relay power scenario, if the destination is used as the jammer, both schemes outperform all existing schemes and achieve the upper bound. In particular, if the SNR is large and the source, the relay, and the cooperative jammer have identical power and channels, both schemes achieve the upper bound for secrecy rate, which is merely $1/2$ bit per channel use lower than the channel capacity without secrecy constraints. We also prove that one of our schemes achieves a positive secrecy rate in the external eavesdropper case in which the relay is trusted and there exists an external eavesdropper.

\end{abstract}


%

\IEEEpeerreviewmaketitle

\section{Introduction}

\IEEEPARstart{I}{nformation}
theoretic security is a problem considered by Shannon whereby no message can be retrieved even if an eavesdropper knows the coding scheme and has infinite computation capabilities \cite{shannon}.
This concept has been well studied for many channels, e.g., the wire-tap channel \cite{wyner}. 
The concept ``secrecy rate'' is proposed in \cite{wyner} for the rate of the communication under the constraint that the information leaked to the eavesdropper per channel use tends to zero when the number of channel uses tends to infinity (a constraint also known as weak secrecy).

As many other classic channels, the secure transmission problem on a two-hop channel with an untrusted relay has been well studied. This channel consists of a pair of source and destination using an untrusted relay to forward the message. The relay is considered to be malicious but cooperative, it overhears the message but makes no change on it. This channel was first studied in \cite{oohama}, in which a rather pessimistic conclusion is drawn that no positive secrecy rate can be achieved by the straightforward transmission scheme. However, a later study in \cite{he} proposed a cooperative jamming \cite{tekin} based approach to achieve a positive secrecy rate, in which a cooperative node (sometimes the destination) is introduced to simultaneously transmit a jamming signal to confuse the relay while the source is transmitting. The relay then encodes its reception and transmits it to the destination. With prior knowledge of the jamming signal, the destination is able to decode the source message.

Several secure transmission schemes have been proposed based on cooperative jamming in \cite{he} and \cite{sun}-\cite{vatedka}. In \cite{he}, the source is encoded with a Gaussian code, the cooperative jammer transmits a random Gaussian signal, and the relay forwards the description of its received signal using the compress-and-forward scheme \cite{elgamal}. A similar scheme based on amplify-and-forward \cite{gastpar} is used in \cite{sun}. The amplify-and-forward based scheme is improved in \cite{zhang} by using a lattice code instead of Gaussian code at the source. This scheme is called modulo-and-forward, since the relay can take a modulo operation w.r.t. lattice structure of the code to achieve a higher secrecy rate. A compute-and-forward \cite{nazer} based scheme was introduced in \cite{he2} for a symmetric two-hop channel, in which both of the transmitting and the jamming messages are encoded with lattice codes. The relay decodes a linear combination of these messages and then sends it to the destination. Albeit the achievable secrecy rate of \cite{he2} is lower than \cite{he}, this compute-and-forward based scheme can be used in a line network since it does not suffer from noise accumulation. In \cite{he3}, a similar compute-and-forward based scheme was introduced which achieves strong secrecy with the same secrecy rate as \cite{he2}. In \cite{richter2}, another compute-and-forward based scheme was proposed for the multi-way relay channel which achieves weak secrecy rate. Bi-directional transmission on this channel is studied in \cite{vatedka}, in which a higher level of secrecy, namely perfect secrecy, is achieved by another compute-and-forward based scheme.


In this paper, we propose two novel secure and reliable transmission schemes based on a modified version of compute-and-forward \cite{jingge}, which we call scaled compute-and-forward. The main contributions of this paper are the following:

\begin{itemize}
\item Two novel scaled compute-and-forward based secure transmission schemes are proposed for the two-hop channel with an untrusted relay, which use a random binning code and a lattice chain to create randomness at the source, respectively. These are the first secure transmission schemes for this problem that are based on the scaled compute-and-forward technique.
\item In the symmetric case where the power and channel gains for the source, the relay, and the destination are identical, both of our schemes achieve a secrecy rate of $1/2\log_2(1/2+{\rm SNR})-1/2$, in which SNR stands for Signal-to-Noise-Ratio. This is merely $1/2$ bit per channel use away from the transmit rate using compute-and-forward on this channel. This rate is upper bound achieving when the SNR is high and is the best secrecy rate achieved so far on this channel.
\item Our schemes significantly improve the achievable secrecy rate and achieve the upper bound in many asymmetric scenarios. In general, our schemes have better performance than other existing schemes in the high SNR scenario for almost all channel configurations.
\item We consider a novel secure transmission problem on the two-hop channel, in which the relay is trusted and there exist an external eavesdropper. The problem is different from the wire-tap type of problem and the untrusted relay problem, and the existing secure transmission schemes cannot be directly used. We prove that one of our schemes can also be applied on this channel and achieves a positive secrecy rate.
\end{itemize}


The paper is organized as follows. In Section~\ref{s:nm} we build up the model, give the state-of-the-art on the problem, and briefly introduce the scaled compute-and-forward technique as proposed in \cite{jingge}. In Section~\ref{s:ts} we introduce a reliable scaled compute-and-forward based code for transmission, which will be used as the transmission code throughout this paper. In Section~\ref{s:scs} we introduce two secure coding schemes which are built upon our reliable transmission scheme and provide secrecy. In Section~\ref{s:comp} we compare the rates of our schemes with the state-of-the-art. In Section~\ref{s:ee}, we consider another two-hop channel model in which the relay is trusted and there exist an external eavesdropper. We show that one of our schemes can also achieve a positive secrecy rate in this case. In Section~\ref{s:con}, we conclude this paper.
%
%
%
%



\section{Preliminaries}\label{s:nm}

\subsection{Model}\label{s:model}

In this paper, we consider the model used in \cite{he}. The model consists of a two-hop channel, in which node $A$ wants to transmit information to node $C$ using an untrusted relay node $R$ to forward the information. To guarantee secure communication, another node $B$, a {\em Cooperative Jammer}, is added to the network, which transmits a jamming signal to confuse the relay. We assume that the communication takes places over two phases, each including $N$ channel uses. 
We use $X^A, X^B, X^R \in \mathbb{R}^N$ for the transmitted sequences of node $A$, the cooperative jammer $B$, and the relay $R$, respectively. We use $Y^R, Y^C_1, Y^C_2\in \mathbb{R}^N$ for the receptions of the relay and node $C$ in Phase 1 and 2, respectively. In the first phase, node $A$ transmits to the relay and the cooperative jammer $B$ simultaneously transmits a jamming signal to confuse the relay. The jamming signal transmitted by the cooperative jammer $B$ is also received by node $C$. We have
\begin{eqnarray}
Y^R&=&X^A+X^B+Z^R_1, \label{eq:yr}\\
Y^C_1&=&X^B+Z^C_1, 
\end{eqnarray}
where $Z^R_1$ and $Z^C_1$ are $N$-dimensional independent Gaussian noise vectors with variance 1 and $\sigma^2$ in each dimension, respectively. Note that when $\sigma=0$, the model is equivalent to the model in which the destination is used as jammer.

In the second phase, the relay transmits to node $C$. We have
\begin{eqnarray}\label{eq:yb}
Y^C_2&=&X^R+Z_2^C,
\end{eqnarray}
where $Z_2^C$ is an $N$-dimensional independent Gaussian noise vector with variance $1$ in each dimension.
The power constraints for the three nodes are defined as
\begin{equation}\label{eq:power}
E[||X^i||^2] \leq NP_i, i \in \{A,B,R\}.
\end{equation}

It seems that we lose some generality by assuming the channel coefficients and $Z^R_1, Z^C_2$ to be unit. However, this assumption is actually w.l.o.g. and can be easily extended to any configuration of power constraints, channel coefficients, and noise variances with the same SNR for the receptions $Y^R$, $Y^C_1$, and $Y^C_2$.

We assume that the power constraints as well as $\sigma^2$ are revealed to all nodes. The source message of node $A$ is defined as $W^A\sim {\cal U} (\{1, 2, \ldots, 2^{NR_s}\})$, where the notation $X \sim {\cal U}({\cal S})$ is used for a random variable $X$ that is uniformly chosen at random from the alphabet ${\cal S}$. A secrecy rate $R_s$ is said to be achievable if for any $\delta>0$, there exists a sequence of $(2^{NR_s}, N)$ codes such that the reliability constraint
\begin{equation}\label{eq:reli}
\lim_{N\to \infty} \Pr(\hat{W}^A \ne W^A)=0
\end{equation}
and the (weak) secrecy constraint
\begin{equation}\label{eq:sec}
\lim_{N \to \infty}\frac{1}{N}I(W^A;Y^R)\leq \delta
\end{equation}
hold.
Here, $\hat{W}^A$ is the estimation of $W^A$ based on the reception $Y^C_1$ and $Y^C_2$ at node $C$.
Further, we use the notation $\mathbb{R}^+$ for the set of positive real numbers, $\mathbb{Z}^+$ for the set of positive integers, and $C(x)$ for the capacity of Gaussian channel with SNR equal to $x$, i.e.,
\begin{equation}
C(x)=\frac{1}{2}\log_2(1+x).
\end{equation}

\begin{figure}[t!]
\centering
\tikzstyle{vertex}=[circle,minimum size=30pt, inner sep=0pt, draw]
\tikzstyle{edge}=[-latex, black, thick]
\tikzstyle{dot}=[edge, dotted]
\tikzstyle{dash}=[edge, densely dashed]
\tikzstyle{loosedo}=[edge, loosely dotted]
\begin{tikzpicture}[scale=2.8]
\node[vertex] (t1) at (0,2.4)  {$A$} ;
\node[vertex] (r) at (1,2.4) {$R$};
\node[vertex] (t2) at (2,2.4) {$C$} ;
\node[vertex] (j) at (1.5,3.2) {$B$};

\draw[edge, above] (0.2,2.7) to node {Phase 1} (0.8, 2.7);
\draw[dot, above](1.35,2.9) to node {} (1.1, 2.7);
\draw[dot, above](1.65,2.9) to node {} (1.9, 2.7);
\draw[edge, above](1.2, 2.1) to node {Phase 2} (1.8, 2.1);

\node (jammer) at (1.5, 2.65) {Phase 1};

\node (xa) at (0, 2.7) {$X^A$};
\node (yr) at (1, 2.7) {$Y^R$};
\node (xb) at (1.5, 2.9) {$X^B$};
\node (xr) at (1, 2.1) {$X^R$};
\node (yc2) at (2, 2.1) {$Y^C_2$};
\node (yc1) at (2, 2.7) {$Y^C_1$};
\end{tikzpicture}
\caption{Two-hop channel with a cooperative jammer.}\label{fig:thc}
\end{figure}
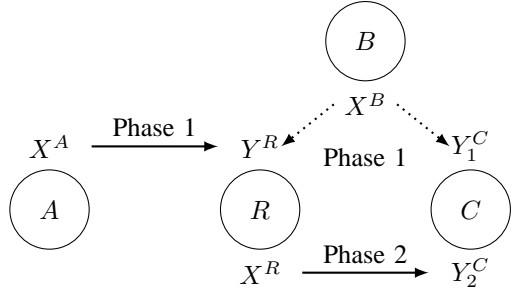

\subsection{State-of-the-Art}

\subsubsection{An Upper Bound on the Secrecy Rate}
An upper bound on the secrecy rate is derived in \cite{he} by transforming this model to an equivalent multiple access wire-tap channel. The secrecy rate is upper bounded by
\begin{equation}\label{eq:ubsec}
R_b=\frac{1}{2}\log_2\frac{(P_A+1)(P_A+P_B+1)-(P_A+\rho)^2}{(P_A+P_B+1)(1-\rho^2)},
\end{equation}
where
\begin{equation}
\rho=\frac{2P_A+P_AP_B+P_B-\sqrt{4P_BP_A^2+4P_BP_A+P_B^2P_A^2+2P_B^2P_A+P_B^2}}{2P_A}.
\end{equation}

\subsubsection{Amplify-and-Forward Based Scheme}

A straightforward amplify-and-forward based scheme is proposed in \cite{sun} for the case that the destination is used as the jammer, i.e., $\sigma=0$. In this scheme the destination transmits a Gaussian jamming signal and the relay simply amplifies the received signal and transmits it to the destination. This scheme achieves any secrecy rate satisfying
\begin{eqnarray}
R_s &<& \frac{1}{2}\log_2\left(1+\frac{P_AP_R}{P_A+P_B+P_R+1}\right)  -\frac{1}{2}\log_2\left(1+\frac{P_A}{P_B+1}\right). \label{eq:sun}
\end{eqnarray}

\subsubsection{Modulo-and-Forward Based Scheme}
In \cite{zhang} another scheme is proposed which uses a lattice code based coding scheme to transmit the message with an extra random vector. The destination, which is also the cooperative jammer ($\sigma=0$), transmits a Gaussian signal to confuse the relay. Due to the lattice chain structure, the relay can take a modulo operation to remove the random part of the transmission which is useless in the decoding at the destination. This results in a higher SNR for the actual message vector. In other words, it is an advanced amplify-and-forward scheme which makes uses of the properties of lattice code and lattice chain. Any secrecy rate satisfying
\begin{equation}\label{eq:zhang}
R_s < \frac{1}{2}\log_2\frac{P_A+P_R+P_AP_R+1}{P_A+P_R+2}-\frac{1}{2}\log_2\left(1+\frac{P_A}{P_B+1}\right)
\end{equation}
is achievable.

\subsubsection{Compress-and-Forward Based Scheme}
A compress-and-forward based scheme is given in \cite{he}, where the relay forwards a description of its noisy reception to the destination. Any secrecy rate
\begin{eqnarray}
R_s < \max_{p_A\leq P_A,p_B \leq P_B}\left(C(\frac{p_A}{(1+\sigma^2+\sigma_c^2)-\sigma^4/(p_B+\sigma^2)})-C(\frac{p_A}{1+p_B})\right) \label{eq:he1}
\end{eqnarray}
is achievable, where 
\begin{equation}
\sigma_c^2=\frac{p_A+1+p_B\sigma^2/(p_B+\sigma^2)}{P_R}.
\end{equation}

\subsubsection{Compute-and-Forward Based Scheme}

Compute-and-forward is a relaying technique proposed in \cite{nazer} in which the relay uses the superimposed nature of Gaussian additive channel and computes the linear combination of the transmitted messages of multiple users instead of individual messages. A compute-and-forward based scheme is proposed in \cite{he2} for a symmetric two-hop channel with jammer and destination collocated, in which node $A$ transmits the source message encoded with a lattice codebook and node $B$ (node $C$) transmits a random codeword choosing uniformly at random from the same lattice codebook. An algebraic proof is given that the sum of two $N$-dimensional lattice codewords will leak no more than $N$ bits of information to the relay. Then, a random binning based scheme is used to eliminate the information leakage. For $P_A=P_B=P_R$ and $\sigma=0$, any secrecy rate satisfying
\begin{equation}\label{eq:he2}
R_s' < \frac{1}{2}\log_2\left(\frac{1}{2}+P_A\right)-1
\end{equation}
is achievable. It is proved in \cite{he3} that this rate is also achievable if we change the weak secrecy constraint (\ref{eq:sec}) to a {\em strong secrecy} constraint
\begin{equation}\label{eq:ssec}
\lim_{N \to \infty}I(W^A;Y^R)\leq \delta
\end{equation}
by replacing the random binning based scheme to a universal hashing function based scheme.

Another compute-and-forward based scheme is proposed in \cite{vatedka}, which also considers the case of $P_A=P_B=P_R$ and $\sigma=0$. The focus of \cite{vatedka} is on {\em ``perfect secrecy''}, which is defined through
\begin{equation}\label{eq:psec}
\lim_{N \to \infty}I(W^A;Y^R)=0.
\end{equation}
A binning approach within the lattice codebook used for both $A$ and $B$ is used. The bins are selected such that for each source message, $A$ randomly selects from a certain bin of codewords with a certain probability mass function. It is proved that if the bins and the probability mass functions are chosen appropriately, perfect secrecy is achievable with any secrecy rate satisfying
\begin{equation}
R_s'' < \frac{1}{2}\log_2\left(\frac{1}{2}+P_A\right)-1-\log_2e.
\end{equation}

This scheme is extended in \cite{vatedka2}, in which the asymmetric channel/power case is considered. It is proved that perfect secrecy is achievable for some asymmetric configurations.

\subsection{Scaled Compute-and-Forward}\label{ss:scf}
Scaled Compute-and-Forward (SCF) as proposed in \cite{jingge} is a generalized version of the traditional compute-and-forward in \cite{nazer}. It allows the senders to scale their lattice codebooks according to their prior knowledge of the channel states to achieve higher computation rates. Here we briefly introduce this technique for a two user Multiple Access Channel (MAC) case.

A lattice $\Lambda$ is a discrete subgroup of $\mathbb{R}^N$ with the property that if $T_1, T_2 \in \Lambda$, then $T_1+T_2 \in \Lambda$. The lattice quantizer $Q_\Lambda$ is defined as $Q_\Lambda(X)=\argmin_{T\in \Lambda}||T- X||$. The fundamental Voronoi region of the lattice is defined as ${\cal V}= \{X \in \mathbb{R}^N| Q_\Lambda(X)=0\}$. The modulo operation is defined as $[X] \Mod{\Lambda}=X-Q_{\Lambda}(X)$. The lattice $\Lambda'$ is said to be nested in $\Lambda$ if $\Lambda' \subseteq \Lambda$. More details of lattices and lattice codes can be found in \cite{erez}.

We consider the two users having power $P_A$ and $P_B$. Firstly, we construct a lattice $\Lambda$. Then we construct two coarse lattices $\Lambda^A, \Lambda^B \subseteq \Lambda$ with second moment $\frac{1}{N {\rm Vol}({\cal V}^i)} \int_{{\cal V}^i} ||X|| dX=\beta_i^2 P_i$, $i \in \{A,B\}$, where ${\cal V}^i$ is the fundamental Voronoi region of $\Lambda^i$ and $\beta_i \in \mathbb{R}^+$ is called the scaling coefficient. Here we assume that $\Lambda, \Lambda^A$ and $\Lambda^B$ are simultaneously good for quantizing and shaping as discussed in \cite{erez}.

For user $i \in\{A, B\}$, we construct the codebook ${\cal L}^i=\Lambda \cap {\cal V}^i$, where ${\cal V}^i$ is the fundamental Voronoi region of $\Lambda^i$. 
User $i$ encodes its message into codeword $T^i$ using the codebook ${\cal L}^i$, and the channel input is formed as
$X^i=[T^i/\beta_i+D^i] \Mod{\Lambda^i / \beta_i}$,
where $D^i\sim{\cal U}({\cal V}^i_C/ \beta_i)$ is called {\em dither}. Clearly, $X^i$ is also uniform in ${\cal V}^i / \beta_i$ and thus it has average power $P_i$. 

The receiver uses the fine lattice $\Lambda$ for decoding the linear sum $a_1 T^A+a_2 T^B, a_1, a_2 \in \mathbb{Z}$. It is proved in \cite{jingge} that the destination is able to reliably decode this linear sum as long as the transmit rates are smaller than the computation rates $R_{\rm CF}^i({\bf a}, {\boldsymbol\beta})$ defined as
\begin{equation}\label{eq:comr2}
R_{\rm CF}^i({\bf a}, {\boldsymbol\beta})=\frac{1}{2}\log_2(\frac{\beta_i^2 P_i}{{\cal N}({\bf a}, {\boldsymbol\beta})}),
\end{equation}
where 
\begin{equation}\label{eq:nabeta}
{\cal N}({\bf a}, {\boldsymbol\beta})=\frac{P_AP_B(a_1\beta_A-a_2\beta_B)^2+(a_1\beta_A)^2P_A+(a_2\beta_B)^2P_B}{P_A+P_B+1},
\end{equation}
${\bf a}=(a_1, a_2)$, and $\boldsymbol\beta=(\beta_A, \beta_B)$.

\remark{For any ${\bf a}$ and ${\boldsymbol\beta}$ it can be derived from (\ref{eq:comr2}) that the computation rates satisfy
\begin{equation}\label{eq:rab}
R_{\rm CF}^A({\bf a}, {\boldsymbol\beta}) \leq C(P_A), R_{\rm CF}^B({\bf a}, {\boldsymbol\beta}) \leq C(P_B).
\end{equation}
}\label{rm:1}



\section{A Scaled Compute-and-Forward Based Code for Reliable Transmission}\label{s:ts}

As introduced in Subsection~\ref{ss:scf}, the relay node is able to compute a linear combination with coefficients ${\bf a}$ by using lattice codebooks with average power $P_i, i\in\{A,B\}$ and scaled with coefficients $\boldsymbol\beta$ if the transmit rates are smaller than the computation rates in (\ref{eq:comr2}). In this section, we will propose a reliable code for our channel, namely an $({\bf a}, {\boldsymbol\beta})$ SCF code. For given power $P_i, i\in{A,B,R}$, this code will guarantee a reliable transmission from $A$ to $C$ for a source symbol chosen uniformly at random from $\{1,2,\ldots, 2^{NR_t^A({\bf a}, {\boldsymbol\beta})}\}$ if
\begin{equation}\label{eq:rtascf}
R_t^A({\bf a}, {\boldsymbol\beta}) < \min (C(P_R), R_{\rm CF}^A({\bf a}, {\boldsymbol\beta}))
\end{equation}
and
\begin{equation}\label{eq:1c}
R_{\rm CF}^B({\bf a}, {\boldsymbol\beta})\leq C(P_B/\sigma^2).
\end{equation}

We will firstly introduce the lattice codebook construction in detail, then describe the transmission process. In the end, we will calculate the rate of information leaked to the relay with this scheme during the transmission.


\subsection{Codebook Construction}\label{ss:cc}

Here we describe our codebook constructed with the SCF technique. For an $({\bf a}, {\boldsymbol\beta})$ SCF code and an arbitrarily chosen positive real number $\delta'$, we select a fine lattice $\Lambda$ and a pair of shaping lattices $\Lambda_C^i({\bf a}, {\boldsymbol\beta}) \subseteq \Lambda, i\in\{A,B\}$ which have the following properties: 
\begin{itemize}
\item {\bf Power:}
We have
\begin{equation}\label{eq:lp}
\begin{array}{l}
\frac{1}{N {\rm Vol}({\cal V}_C^i({\bf a}, {\boldsymbol\beta}))}\int_{{\cal V}_C^i({\bf a}, {\boldsymbol\beta})} ||X||^2 dX=\beta_i^2 P_i, i\in\{A, B\},
\end{array}
\end{equation}
where ${\cal V}_C^i$ is the fundamental Voronoi region of $\Lambda_C^i({\bf a}, {\boldsymbol\beta})$.
\item{\bf Nesting:}
The coarser one of $\Lambda_C^i({\bf a}, {\boldsymbol\beta})$ is nested in the finer one.
\item{\bf Rate:}
Denote $\hat{R}_t^i({\bf a}, {\boldsymbol\beta}) = \frac{1}{N} \log_2 |\Lambda \cap {\cal V}_C^i({\bf a}, {\boldsymbol\beta})|$. Then, for the chosen $\delta'$, we have
\begin{equation}\label{eq:rtcf}
R_{\rm CF}^i({\bf a}, {\boldsymbol\beta}) - \delta' < \hat{R}_t^i({\bf a}, {\boldsymbol\beta})<R_{\rm CF}^i({\bf a}, {\boldsymbol\beta}).
\end{equation}
\item{\bf Goodness:}
These lattices are all good in both quantizing and shaping in the sense of \cite{erez}. 
\end{itemize}

By \cite{erez}, we can find lattices satisfying the above-mentioned properties. Then, we construct the lattice codebooks $\Lambda \cap {\cal V}_C^i({\bf a}, {\boldsymbol\beta})$ for transmission. 


\subsection{Reliable Transmission Process}\label{ss:rt}

\begin{itemize}


\item {\bf Phase 1, node $A$.}

Firstly, the message $W^A$ is uniquely mapped to a lattice vector in $\Lambda \cap {\cal V}_C^A({\bf a}, {\boldsymbol\beta})$ by the encoder, Then, a dither $D^A$ is uniformly chosen from the scaled Voronoi region ${\cal V}^A_C({\bf a}, {\boldsymbol\beta})/\beta_A$. Note that dithers are chosen to fulfill the power constraints of lattice codes and are revealed to all nodes. The transmitted lattice vector of node $A$ is 
\begin{equation}\label{eq:xa}
X^A = [T^A/\beta_A+D^A] \Mod{\Lambda_C^A({\bf a}, {\boldsymbol\beta})/\beta_A}.
\end{equation}

\item {\bf Phase 1, node $B$.}

Node $B$ transmits a jamming signal, namely $V^B$, which is uniformly chosen at random from $\Lambda \cap {\cal V}_C^B({\bf a}, {\boldsymbol\beta})$.
The transmitted vector is thus
\begin{equation}\label{eq:xb}
X^B=[V^B/\beta_B+D^B] \Mod{\Lambda_C^B({\bf a}, {\boldsymbol\beta})/\beta_B},
\end{equation}
in which $D^B \sim {\cal U}({\cal V}^B_C({\bf a}, {\boldsymbol\beta})/\beta_B)$. 
It it clear from the definition that the average power of both $X^A$ and $X^B$ does not exceed the power constraint of (\ref{eq:power}).

\item {\bf Phase 1, node $C$}

By our codebook construction, node $C$ can reliably decode $V^B$ if (\ref{eq:1c}) holds.

\item {\bf Phase 1, the relay.}

Upon receiving $Y^R$ in (\ref{eq:yr}), by \cite{jingge}, the relay is able to decode $$U^R=a_1T^A+a_2V^B$$ with the lattice $\Lambda$ with high probability if 
\begin{equation}\label{eq:racf}
\hat{R}^i_t({\bf a}, {\boldsymbol\beta}) < R_{\rm CF}^i({\bf a}, {\boldsymbol\beta}), i\in\{A, B\},
\end{equation}
which has already be guaranteed by the codebook construction in (\ref{eq:rtcf}).

\item{\bf Phase 2, the relay}

The relay firstly scales the decoded vector down by computing $U^R/a_1 =T^A+(a_2/a_1)V^B$. Then, similar to the compute-and-forward scheme proposed for the two-way relay channel \cite{nazer2}, a modulo operation is taken on the decoded vector. The lattice for the modulo operation $\Lambda^*$ should be chosen such that $\Lambda^* \subseteq \Lambda_C^A({\bf a}, {\boldsymbol\beta})$ to guarantee that the vector $T^A$ can be retrieved by node $C$. For the sake of power, we let the relay take a modulo operation on the decoded vector w.r.t. $\Lambda_C^A({\bf a}, {\boldsymbol\beta})$. We denote the resulting vector as $\tilde{U}^R$ and 
\begin{equation}
\tilde{U}^R =U^R/a_1 \Mod{\Lambda_C^A({\bf a}, {\boldsymbol\beta})}=(T^A+(a_2/a_1)V^B) \Mod{\Lambda_C^A({\bf a}, {\boldsymbol\beta})}.
\end{equation}
Then, the relay transmits this vector using any capacity achieving channel code on the Additive White Gaussian Noise (AWGN) channel. By definition, the entropy of this vector has the property of
\begin{equation}\label{eq:hur}
\frac{1}{N} H(\tilde{U}^R) \leq \hat{R}_t^A({\bf a}, {\boldsymbol\beta})\leq R_{\rm CF}^A({\bf a}, {\boldsymbol\beta}) \leq C(P_A).
\end{equation} 
When $P_R \geq P_A$, this vector can be reliably transmitted by the relay straightforwardly.
When $P_R < P_A$, we consider a long term of transmission during which the model is used for $K \in \mathbb{Z}^+$ times. We only use $\lceil K \frac{C(P_R)}{\hat{R}_t^A({\bf a}, {\boldsymbol\beta})}\rceil$ times Phase 1 and fully use all $K$ times Phase 2 of these model uses. By choosing $K$ sufficiently large, the transmit rate can be made arbitrarily close to
\begin{equation}\label{eq:rta0}
R_t^A({\bf a}, {\boldsymbol\beta}) \frac{C(P_R)}{\hat{R}_t^A({\bf a}, {\boldsymbol\beta})}.
\end{equation}
Combining these two cases, in a long term transmission, any rate satisfying (\ref{eq:rtascf}) is achievable.

\item {\bf Phase 2, node $C$.}

Since node $C$ can reliably decode $\tilde{U}^R$, it can then decode $T^A$ by computing
\begin{eqnarray}
\lefteqn{[\tilde{U}^R -(a_2/a_1)V^B] \Mod{\Lambda_C^A({\bf a}, {\boldsymbol\beta})}} \nonumber \\
&=& [(T^A+(a_2/a_1)V^B) \Mod{\Lambda_C^A({\bf a}, {\boldsymbol\beta})} -(a_2/a_1)V^B]  \Mod{\Lambda_C^A({\bf a}, {\boldsymbol\beta})}  \nonumber \\
&=&[T^A+(a_2/a_1)V^B-(a_2/a_1)V^B] \Mod{\Lambda_C^A({\bf a}, {\boldsymbol\beta})} \nonumber \\
&= & [T^A] \Mod{\Lambda_C^A({\bf a}, {\boldsymbol\beta})} \nonumber \\
&=& T^A.
\end{eqnarray}
Since $T^A$ is reliably decoded, $W^A$ can then be retrieved.
\end{itemize}

As we have already discussed the reliability of the decoding in each step of the process, by choosing $\delta'$ arbitrarily small we have the following lemma.

\lemma{For any ${\bf a}, {\boldsymbol\beta}$, an $({\bf a}, {\boldsymbol\beta})$ SCF code guarantees power constraint (\ref{eq:power}) and reliability constraint (\ref{eq:reli}) with any rate satisfying (\ref{eq:rtascf}) if (\ref{eq:1c}) holds.}\label{lm:scf}

\subsection{Information Leakage Rate}

By Lemma~\ref{lm:scf}, we guarantee that the information can be reliably transmitted from the source to the destination with the given power constraint. However, during the process, part of the information is leaked to the relay. Here we define the information leakage rate $R_o({\bf a}, {\boldsymbol\beta})$ (sometimes referred as equivocation rate) as
\begin{equation}
R_o({\bf a}, {\boldsymbol\beta}) = \frac{1}{N}I(W^A;Y^R)
\end{equation}
and bound it by
\begin{eqnarray}
R_o({\bf a}, {\boldsymbol\beta}) &=&\frac{1}{N}(H(W^A)-H(X^A,X^B|Y^R) -H(W^A|Y^R,X^A,X^B)+H(X^A,X^B|Y^R,W^A)) \nonumber \\
&=& \frac{1}{N}(H(W^A)-H(X^A, X^B)+I(X^A,X^B;Y^R)) \nonumber\\
&<& \frac{1}{N}(H(W^A)-H(X^A)-H(X^B)) +C(P_A+P_B) \nonumber \\
&=& R_t^A({\bf a}, {\boldsymbol\beta}) -R_t^A({\bf a}, {\boldsymbol\beta}) -R_t^B({\bf a}, {\boldsymbol\beta}) +C(P_A+P_B) \nonumber \\
&=& -R_t^B({\bf a}, {\boldsymbol\beta})+C(P_A+P_B). \label{eq:il4}
\end{eqnarray} 

The second equality holds since the third term on the RHS of the first equality is 0 since $W^A$ is one-to-one mapped to $X^A$. Further, the fourth term is also 0 for that the relay knows $X^A$ by knowing $W^A$, and it can reliably decode $X^B$ when $X^A$ is known. Then, the inequality follows from the capacity for Gaussian MAC and the third equality follows from the definition of the $({\bf a}, {\boldsymbol\beta})$ SCF code.
In the next section, we will propose two schemes to eliminate the information leakage.

\section{Secure Coding Schemes}\label{s:scs}

In the previous section, we have proposed a reliable code for transmission, namely an $({\bf a}, {\boldsymbol\beta})$ SCF code. For any ${\bf a}, {\boldsymbol\beta}$, a reliable transmission of a source symbol chosen uniformly at random from $\{1,2,$ $\ldots, 2^{NR_t^A({\bf a}, {\boldsymbol\beta})}\}$ is guaranteed if conditions (\ref{eq:rtascf}) and (\ref{eq:1c}) hold. Then, we bounded the information leakage rate during the process $R_o({\bf a}, {\boldsymbol\beta})$ in (\ref{eq:il4}).

In this section, we introduce two schemes of adding extra randomness at the source, which can eliminate the information leakage. Both schemes are built upon the $({\bf a}, {\boldsymbol\beta})$ SCF code. The first one uses the classical random binning idea and is constructed in a two-layer structure. We use the $({\bf a}, {\boldsymbol\beta})$ SCF code as inner code and a random binning code as outer code. The second scheme is a lattice chain based scheme using an $({\bf a}, {\boldsymbol\beta})$ SCF lattice chain code, in which a mid-layer lattice is added to the lattice codebook of the $({\bf a}, {\boldsymbol\beta})$ SCF code to create randomness. Since this code is a modified version of $({\bf a}, {\boldsymbol\beta})$ SCF code, we will describe the difference between this code and the $({\bf a}, {\boldsymbol\beta})$ SCF code described in Section~\ref{s:ts}. 


\subsection{Random Binning Based Scheme}\label{ss:rbbs}

The classical random binning idea is introduced in \cite{wyner} and widely used in many secure transmission scenarios. Here, we borrow the idea of the random binning codes from \cite{csiszar} and the two-layer structure from \cite{he2}. We propose a random binning based scheme (RB scheme), which is constructed by an $({\bf a}, {\boldsymbol\beta})$ SCF code as inner code and a random binning code as outer code. The random binning code is designed to encode the messages into a long sequence of lattice codewords of a chosen $({\bf a}, {\boldsymbol\beta})$ SCF code. Here we introduce our random binning code in detail.

\subsubsection{Random Binning Code}

\begin{itemize}
\item {\bf Codebook Construction} Generate $2^{\lfloor l H(W^A) \rfloor}$ bins, where $l$ should be chosen sufficiently large. Label each bin by a different length-$l$ typical sequence of $W^A$. The size of each bin (the upper bound for the number of the codewords in each bin) is $2^{\lfloor l' N R_o({\bf a}, {\boldsymbol\beta})\rfloor}$, where $l'=  l\frac{H(W^A)}{N(R_t^A({\bf a}, {\boldsymbol\beta})-R_o({\bf a}, {\boldsymbol\beta}))}$.

Generate $2^{NR_t^A({\bf a}, {\boldsymbol\beta})\lceil l'\rceil}$ codewords. The codewords are length $\lceil l' \rceil$ sequences of $N$-dimensional lattice codewords generated with the codebook described in Section~\ref{ss:cc}.
Put the codewords into the bins uniformly at random until all bins are filled.

\item {\bf Encoding} For each length-$l$ sequence of source messages, the encoder chooses the bin with the same label, then chooses a codeword from the bin uniformly at random and transmits it. If the message sequence does not match any label of bins, or there is no codeword in the matching bin, it transmits a random length-$\lceil l' \rceil$ sequence of lattice codewords as its codeword. The code structure of the RB scheme is illustrated in Fig.~\ref{fig:encode}.

\item {\bf Transmission} Here, we consider that we use our model $\lceil l'\rceil$ times. Each time a lattice codeword is reliably transmitted from node $A$ to $B$. Thus, after $\lceil l'\rceil$ times, a random binning codeword is reliably transmitted. For each phase, the channel is used for $N\lceil l'\rceil$ times.

\item {\bf Decoding} By receiving the codeword, it looks that up into the codebook and uses the label of the bin as the estimation.

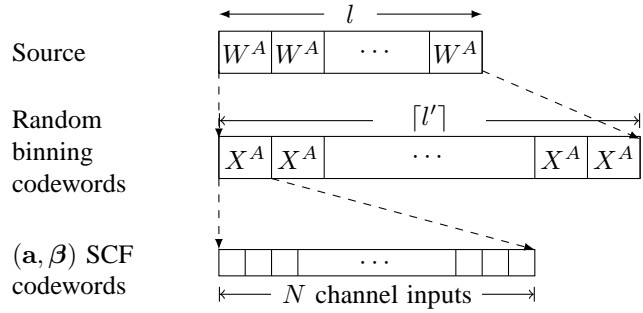
\begin{figure}
\centering

\begin{tikzpicture}[scale=.35]
\tikzstyle{->}=[-latex];

\node [text width = 20mm] at (-15,2) {Source};

\node [text width = 20mm] at (-15,-1.5) {\singlespace Random binning codewords};

\node [text width = 20mm] at (-15, -6) {\singlespace $({\bf a}, {\boldsymbol\beta})$ SCF codewords};

\begin{scope}[xshift=-6cm,yshift=2cm]
\draw (-4,-.8) rectangle (6,.8);
\foreach \x in {-4,-2,0,4,6}
 {
    \draw (\x,-.8) -- (\x,.8);
}

\node[fill=white,inner sep=0pt] at (-3,0) {$W^A$};
\node[fill=white,inner sep=0pt] at (-1,0) {$W^A$};
\node[fill=white,inner sep=0pt, minimum width=10mm] at (2,0) {$\cdots$};
\node[fill=white,inner sep=0pt] at (5,0) {$W^A$};

\draw[->] (0,1.45) -- (-4,1.45);
\draw[->] (2,1.45) -- (6,1.45);

\node[fill=white,inner sep=0pt] at (1,1.45) {$l$};



\end{scope}

\draw[dashed,->] (-10,1.3) -- (-10,-1.3);
\draw[dashed,->] (0,1.3) -- (6,-1.3);

\begin{scope}[xshift=-6cm,yshift=-2cm]
\draw (-4,-.8) rectangle (12,.8);
\foreach \x in {-4,-2,0,8,10,12}
 {
    \draw (\x,-.8) -- (\x,.8);
}
\node[fill=white,inner sep=1pt] at (-3,0) {$X^A$};
\node[fill=white,inner sep=1pt] at (-1,0) {$X^A$};
\node[fill=white,inner sep=1pt] at (4,0) {$\cdots$};
\node[fill=white,inner sep=1pt] at (9,0) {$X^A$};
\node[fill=white,inner sep=1pt] at (11,0) {$X^A$};




\draw[|<->|] (-4,1.4) -- (12,1.4);
\node[fill=white,inner sep=0pt, minimum width=15mm] at (4,1.5) {$\lceil l' \rceil$};


\end{scope}

\draw[dashed,->] (-10,-2.8) -- (-10,-5.5);
\draw[dashed, ->] (-8,-2.8) -- (2,-5.5);

\begin{scope}[xshift=-6cm,yshift=-6cm]
\draw (-4,-.5) rectangle (8,.5);
\foreach \x in {-3,-2,-1,5,6,7}
 {
    \draw (\x,-.5) -- (\x,.5);
}
\node[fill=white,inner sep=1pt] at (2,0) {$\cdots$};

\draw[|<->|] (-4,-1.2) -- (8,-1.2);
\node[fill=white,inner sep=1pt, minimum width=28mm] at (2,-1.3) {$N$ channel inputs};

\end{scope}

\end{tikzpicture}
\endpgfgraphicnamed
\caption{Structure of the random binning based scheme.}\label{fig:encode}
\end{figure}

\end{itemize}

\subsubsection{Reliability}\label{sss:re}

Since the reliability of the $({\bf a}, {\boldsymbol\beta})$ SCF code is already shown in Section~\ref{s:ts}, we now show the reliability of the random binning code. 

A length-$l$ sequence of source messages can be reliably retrieved if the length-$\lceil l' \rceil$ sequence of lattice codewords is reliably decoded and is the codeword for the correct messages. The former is guaranteed by Lemma~\ref{lm:scf}. An error in the latter can be caused either by an unlabeled message or an empty bin. There are two situations for the unlabeled messages. 
1, the message is typical but there is no matching label. 2, the message is not typical. By the property of typicality, the probability for both situations to occur are negligible when $N,l \to \infty$. Moreover, since the expected number of codewords in one bin is almost $2^{lNR_o({\bf a}, {\boldsymbol\beta})}$, by the law of large number, the probability of the existence of empty bins is also negligible when $N,l \to \infty$. Hence, the estimation error is vanishing when $N,l \to \infty$.

\subsubsection{Information Leakage Rate}

Here, we show that the RB scheme achieves the information theoretic security. 

\lemma{For any $\delta>0$, there exist a sequence of codes constructed with an $({\bf a}, {\boldsymbol\beta})$ SCF code as inner code and a random binning code as outer code which achieves
\begin{equation}
\frac{1}{N}I(W^A;Y^R)<\delta
\end{equation}
}\label{lm:rb}

The proof of this lemma is given in Appendix~\ref{ap:1}.

\subsubsection{Achievable Secrecy Rate}

Here, we discuss three cases of whether the relay has limited power and whether $\sigma$ is larger than a threshold $\underline{\sigma}$ where
\begin{equation}
\underline{\sigma}=\sqrt{1+\frac{1+P_A+P_B}{P_AP_B-P_A-1}}.
\end{equation}
Note that a chosen $({\bf a}, {\boldsymbol\beta})$ SCF code is associated with a threshold for transmit rate $\hat{R}_t^A({\bf a}, {\boldsymbol\beta})$. The actual transmit rate $R_t^A({\bf a}, {\boldsymbol\beta})$ can be chosen arbitrarily in $[0,\hat{R}_t^A({\bf a}, {\boldsymbol\beta})]$. Hence, for each case, we specify the code and transmit rate, i.e., ${\bf a}$, ${\boldsymbol\beta}$, and $R_t^i({\bf a}, {\boldsymbol\beta})$.

\begin{itemize}
\item $P_R \geq P_A$ and $\sigma \leq \underline{\sigma}$.

Firstly, for any $({\bf a}, {\boldsymbol\beta})$, we can bound the achievable secrecy rate of a RB scheme by 
\begin{eqnarray}
R_s&=&\frac{lH(W^A)}{N\lceil l' \rceil}  \nonumber \\
&\geq & R_t^A({\bf a}, {\boldsymbol\beta})-R_o({\bf a}, {\boldsymbol\beta})-\epsilon \nonumber \\
&>& R_t^A({\bf a}, {\boldsymbol\beta})+R_t^B({\bf a}, {\boldsymbol\beta})-C(P_A+P_B)-\epsilon \label{eq:rsl}
\end{eqnarray}
where $\epsilon$ can be made arbitrarily small by choosing sufficiently large $l$ and small $\delta'$. The first and the second inequality simply follow from the definition of $l'$ and $R_o({\bf a}, {\boldsymbol\beta})$, respectively.

Then, when $P_R \geq P_A$ and $\sigma \leq \underline{\sigma}$, (\ref{eq:rsl}) is maximized when $R_t^i ({\bf a}, {\boldsymbol\beta})=\hat{R}_t^i({\bf a}, {\boldsymbol\beta})$. Note that by (\ref{eq:rtcf}), $\hat{R}_t^i({\bf a}, {\boldsymbol\beta})$ is arbitrarily close to $R_{\rm CF}^i({\bf a}, {\boldsymbol\beta})$ when $\delta'$ is chosen arbitrarily small. Hence, any secrecy rate satisfying 
\begin{equation}\label{eq:rbrs1}
R_s < \max_{{\bf a}, {\boldsymbol\beta}}R_s({\bf a}, {\boldsymbol\beta})
\end{equation}
is achievable, where
\begin{equation}
R_s({\bf a}, {\boldsymbol\beta})=R_{\rm CF}^A({\bf a}, {\boldsymbol\beta})+R_{\rm CF}^B({\bf a}, {\boldsymbol\beta})-C(P_A+P_B).
\end{equation}
It can be easily calculated that the maximum is reached when ${\bf a}=(1,1)$ and $\frac{\beta_A}{\beta_B}=\sqrt{\frac{P_B(1+P_A)}{P_A(1+P_B)}}$. In this case we have
\begin{eqnarray}
\max_{{\bf a}, {\boldsymbol\beta}}R_s({\bf a}, {\boldsymbol\beta}) &=&\frac{1}{2}\log_2\frac{1+P_A+P_B}{(\sqrt{(1+P_A)(1+P_B)}-\sqrt{P_AP_B})^2}-1.  \label{eq:rsmax}
\end{eqnarray}

\item $P_R \geq P_A$ and $\sigma > \underline{\sigma}$.

In this case, first of all, if we simply apply the code of the previous case, $R_t^B({\bf a}, {\boldsymbol\beta})$ will be larger than $C(P_B/\sigma^2)$ and the transmitted vector $V^B$ will not be decodable at node $C$. Also, it can be calculated that the achievable secrecy rate is not optimal by adjusting ${\bf a}, {\boldsymbol\beta}$ such that $\hat{R}_t^i({\bf a}, {\boldsymbol\beta})<C(P_B/\sigma^2)$.

Actually, the maximum secrecy rate will be given by choosing ${\bf a}$, ${\boldsymbol\beta}$, and $R_t^i({\bf a}, {\boldsymbol\beta})$ such that $R_{\rm CF}^A({\bf a}, {\boldsymbol\beta})=C(P_A)$, $R_t^A({\bf a}, {\boldsymbol\beta})$ very close to $C(P_A)$, and $R_t^B({\bf a}, {\boldsymbol\beta})$ very close to $C(P_B/\sigma^2)$. The choice of $R_t^B({\bf a}, {\boldsymbol\beta})$ is feasible because $\hat{R}_t^B({\bf a}, {\boldsymbol\beta})$ can be chosen arbitrarily close to $R_{\rm CF}^B({\bf a}, {\boldsymbol\beta})$ and when $\sigma > \underline{\sigma}$, $R_{\rm CF}^B({\bf a}, {\boldsymbol\beta}) > C(P_B/\sigma^2)$. Thus, by (\ref{eq:rsl}), any secrecy rate satisfying
\begin{equation}\label{eq:rbrs2}
R_s < C(P_A)+C(P_B/\sigma^2)-C(P_A+P_B)
\end{equation}
is achievable.

\item $P_R < P_A$. 

In this case, for $\sigma \leq \underline{\sigma}$ and $\sigma > \underline{\sigma}$, the setting for ${\bf a}$, ${\boldsymbol\beta}$, and $R_t^i({\bf a}, {\boldsymbol\beta})$ are identical to the previous two cases, respectively. The difference is that the relay should apply the transmission scheme for $P_R < P_A$, which has already been stated in Subsection~\ref{ss:rt}. As a result, the achievable secrecy rate is simply the achievable secrecy rate of the previous two cases times $\frac{C(P_R)}{\hat{R}_t^A ({\bf a}, {\boldsymbol\beta})}$.

\end{itemize}
Combining the three cases, we have the following theorem.

\theorem{For a two-hop channel with an untrusted relay, with the RB scheme, any secrecy rate $R_s$ satisfying
\begin{equation}\label{eq:rsrb11}
R_s < \max_{{\bf a},{\boldsymbol\beta}}[\min(\frac{C(P_R)}{R_{\rm CF}^A ({\bf a}, {\boldsymbol\beta})},1)R_s({\bf a}, {\boldsymbol\beta})]
\end{equation}
is achievable if $\sigma \leq \underline{\sigma}$ and
\begin{equation}\label{eq:rsrb12}
R_s <\min(\frac{C(P_R)}{C(P_A)},1)(C(P_A)+C(P_B/\sigma^2)-C(P_A+P_B))
\end{equation}
is achievable if $\sigma > \underline{\sigma}$.}\label{lm:rb2}

\subsection{Lattice Chain Based Scheme}


The lattice chain based scheme (LC scheme) is inspired by the lattice chain code used in \cite{richter}. Here, we propose an $({\bf a}, {\boldsymbol\beta})$ SCF lattice chain code, which is an $({\bf a}, {\boldsymbol\beta})$ SCF code with the transmitted lattice vector splitting into two parts, a message vector and a random vector. Now, we describe this code in detail. Since it is modified over an $({\bf a}, {\boldsymbol\beta})$ SCF code, we only focus on the parts that are modified. All the notations and terms have the same meanings as in Section~\ref{s:ts} without further explanation.

\subsubsection{Coding Scheme}

The codebook of an $({\bf a}, {\boldsymbol\beta})$ SCF lattice chain code is also constructed with the lattices $\Lambda$ and $\Lambda_C^i({\bf a}, {\boldsymbol\beta})$ of an $({\bf a}, {\boldsymbol\beta})$ SCF code. Besides, a mid-layer lattice $\Lambda_E^A({\bf a}, {\boldsymbol\beta})$ for which $\Lambda_C^A({\bf a}, {\boldsymbol\beta}) \subseteq \Lambda_E^A({\bf a}, {\boldsymbol\beta}) \subseteq \Lambda$ is introduced for the codebook construction. For arbitrarily chosen $\delta'>0$ and $\delta'' \in (0,(\sum_{i \in \{A, B\}}R^i_{\rm CF}({\bf a}, {\boldsymbol\beta}) - C(P_A+P_B))/2]$, these lattices should satisfy all properties listed in Subsection~\ref{ss:cc}, and three additional properties as follows.
\begin{itemize}
\item{\bf Rate of the Randomness:}
For the given $\delta''$, we have
\begin{equation}\label{eq:macab}
R_o({\bf a}, {\boldsymbol\beta}) - \delta'' < R_e^A({\bf a}, {\boldsymbol\beta})< R_o({\bf a}, {\boldsymbol\beta}),
\end{equation}
where $R_e^A({\bf a}, {\boldsymbol\beta}) = \frac{1}{N} \log_2 |\Lambda_E^A({\bf a}, {\boldsymbol\beta}) \cap {\cal V}_C^A({\bf a}, {\boldsymbol\beta})|$.
\item{\bf Nesting of $\Lambda_E^A({\bf a}, {\boldsymbol\beta})$:}
The coarser one of $\Lambda_E^A({\bf a}, {\boldsymbol\beta})$ and $\Lambda_C^B({\bf a}, {\boldsymbol\beta})$ is nested in the finer one. 
\item{\bf Goodness of $\Lambda_E^A({\bf a}, {\boldsymbol\beta})$:}
The lattice $\Lambda_E^A({\bf a}, {\boldsymbol\beta})$ is good at both quantizing and shaping.
\end{itemize}

By \cite{erez}, lattices satisfying these properties can be found. We then define 
$\hat{R}_s^A({\bf a}, {\boldsymbol\beta}) = \frac{1}{N} \log_2 |\Lambda \cap {\cal V}_E^A({\bf a}, {\boldsymbol\beta})|$ as the rate of the lattice codebook for the messages and
$\hat{R}_t^A({\bf a}, {\boldsymbol\beta})= \frac{1}{N} \log_2 |\Lambda \cap {\cal V}_C^A({\bf a}, {\boldsymbol\beta})|$ as rate of the codebook for transmission.

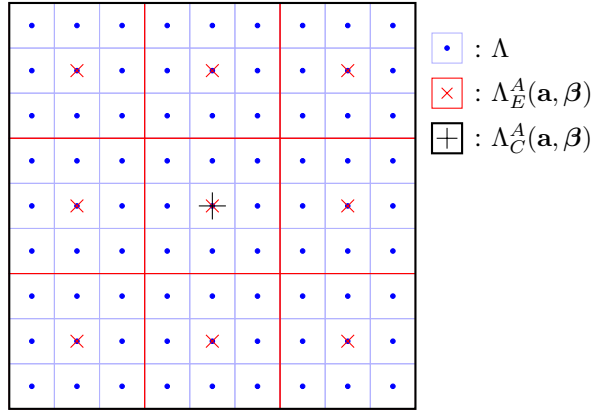
\begin{figure}[!ht]
\centering
\begin{tikzpicture}[scale=0.6]
\tikzstyle{finelattice}=[rectangle,draw=blue!30,inner sep=0pt,minimum size=6mm]
\tikzstyle{finelatticesamp}=[rectangle,draw=blue!30,inner sep=0pt,minimum size=4mm]
\tikzstyle{midlattice}=[rectangle,draw=red,inner sep=0pt,minimum size=18mm]
\tikzstyle{midlatticesamp}=[rectangle,draw=red,inner sep=0pt,minimum size=4mm]
\tikzstyle{coarselattice}=[rectangle,draw=black, thick, inner sep=0pt,minimum size=54mm]
\tikzstyle{coarselatticesamp}=[rectangle,draw=black, thick, inner sep=0pt,minimum size=4mm]

\node at (0,1) {};

\foreach \x in {2,3,...,10}
\foreach \y in {1,...,9}
{
\node at (\x,\y) [finelattice] {};
\draw (\x,\y) plot [mark=*, mark options={color=blue},mark size=0.5mm] coordinates{(\x,\y)} node{};
}
\foreach \x in {3,6,9}
\foreach \y in {2,5,8}
{
\node at (\x,\y) [midlattice] {};
\draw (\x,\y) plot [mark=x, mark options={color=red}, mark size=2mm] coordinates{(\x,\y)} node{};
}
\node at (6,5) [coarselattice] {};
\draw (6,5) plot [mark=+, mark size=3mm] coordinates{(6,5)} node{};

\node at (11.2,8.5) [finelatticesamp] {};
\draw (11.2,8.5) plot [mark=*, mark options={color=blue},mark size=0.5mm] coordinates{(11.2,8.5)} node{};
\draw (11.6,8.5) node[anchor=west] {: $\Lambda$};

\node at (11.2,7.5) [midlatticesamp] {};
\draw (11.2,7.5) plot [mark=x, mark options={color=red}, mark size=1.7mm] coordinates{(11.2,7.5)} node{};
\draw (11.6,7.5)  node[anchor=west] {: $\Lambda_E^A({\bf a},{\boldsymbol\beta})$};

\node at (11.2,6.5) [coarselatticesamp] {};
\draw (11.2,6.5) plot [mark=+, mark size=2.3mm] coordinates{(11.2,6.5)} node{};
\draw (11.6,6.5)  node[anchor=west] {: $\Lambda_C^A({\bf a},{\boldsymbol\beta})$};


\end{tikzpicture}
\caption{A codebook of node $A$ for an $({\bf a}, {\boldsymbol\beta})$ SCF lattice chain code.}\label{fig:lc}
\end{figure}

In Fig.~\ref{fig:lc} we show the structure of a codebook of node $A$.

Clearly, we have
\begin{equation}\label{eq:rta2}
\hat{R}_t^A({\bf a}, {\boldsymbol\beta}) = \hat{R}_s^A({\bf a}, {\boldsymbol\beta}) + R_e^A({\bf a}, {\boldsymbol\beta}).
\end{equation}
The source symbol chosen uniformly at random from $\{1,2,\ldots,2^{NR_s^A({\bf a}, {\boldsymbol\beta})}\}, R_s^A({\bf a}, {\boldsymbol\beta}) \in [0, \hat{R}_s^A({\bf a}, {\boldsymbol\beta})]$ is mapped to a codeword in the lattice codebook $\Lambda \cap {\cal V}_E^A({\bf a}, {\boldsymbol\beta})$. Further, we denote
\begin{equation}\label{eq:rta}
R_t^A({\bf a}, {\boldsymbol\beta}) = R_s^A({\bf a}, {\boldsymbol\beta}) + R_e^A({\bf a}, {\boldsymbol\beta}).
\end{equation}

We assume all the lattices and codebooks are revealed to all four nodes.

\subsubsection{Transmission Process}

The transmission process is similar to the transmission process described in Subsection~\ref{ss:rt}. Here, we only focus on the steps which are different, which are {\bf (Phase 1, node $A$)}, {\bf (Phase 1, the relay)}, {\bf (Phase 2, the relay)}, and {\bf (Phase 2, node $C$)}.


\begin{itemize}
\item {\bf Phase 1, node $A$.}
Firstly, the message $W^A$ is uniquely mapped to a lattice vector in $\Lambda \cap {\cal V}_E^A({\bf a}, {\boldsymbol\beta})$ by the encoder, Then the encoder adds a vector $V^A$ which is chosen uniformly at random from $\Lambda_E^A({\bf a}, {\boldsymbol\beta}) \cap {\cal V}^A_C({\bf a}, {\boldsymbol\beta})$. Then, a dither $D^A \sim {\cal U}({\cal V}^A_C({\bf a}, {\boldsymbol\beta})/\beta_A)$ is chosen. The transmitted lattice vector of node $A$ is 
\begin{equation}\label{eq:xa}
X^A = [(T^A+V^A)/\beta_A+D^A] \Mod{\Lambda_C^A({\bf a}, {\boldsymbol\beta})/\beta_A}.
\end{equation}


\item {\bf Phase 1, the relay.}
Upon receiving $Y^R$ in (\ref{eq:yr}), by \cite{jingge}, the relay can reliably decode $$U^R=a_1(T^A+V^A)+a_2V^B$$ with the lattice $\Lambda$.


\item{\bf Phase 2, the relay}
The relay firstly scales the decoded vector down by computing $U^R/a_1 =(T^A+V^A)+(a_2/a_1)V^B$. Then, instead of $\Lambda_C^A({\bf a}, {\boldsymbol\beta})$, the relay takes a modulo operation on the decoded vector w.r.t. $\Lambda_E^A({\bf a}, {\boldsymbol\beta})$. We denote the resulting vector as $\tilde{U}^{R}_*$ and
\begin{equation}
\tilde{U}^{R}_*=U^R/a_1 \Mod{\Lambda_E^A({\bf a}, {\boldsymbol\beta})}=(T^A+(a_2/a_1)V^B)\Mod{\Lambda_E^A({\bf a}, {\boldsymbol\beta})}.
\end{equation}
The relay then transmits this vector using any capacity achieving channel code on the AWGN channel. Since by definition we have $\frac{1}{N}H(\tilde{U}^{R}_*) \leq \hat{R}_s^A({\bf a}, {\boldsymbol\beta})$, the transmission is reliable when $C(P_R) \geq \hat{R}_s^A({\bf a}, {\boldsymbol\beta})$. Thus we have the power constraint
\begin{equation}\label{eq:hatpa}
P_R > 2^{2\hat{R}_s^A({\bf a}, {\boldsymbol\beta})}-1.
\end{equation}
Note that if $P_R$ is smaller than the requirement in the constraint, a similar approach as the one stated in Section~\ref{s:ts} can be used. However, it can be calculated that the optimal solution is that we adjust ${\bf a}, {\boldsymbol\beta}$ as well as the codebook to $P_R$. The details and the achievable rate of this solution will be given later in this subsection.



\item {\bf Phase 2, Node $C$.}
Since the vector $\tilde{U}^{R}_*$ is reliably decoded, node $C$ can then decode $T^A$ by computing
\begin{eqnarray}
\lefteqn{[\tilde{U}^{R}_* -(a_2/a_1)V^B] \Mod{\Lambda_E^A({\bf a}, {\boldsymbol\beta})}} \nonumber \\
&=& [(T^A+(a_2/a_1)V^B) \Mod{\Lambda_E^A({\bf a}, {\boldsymbol\beta})} -(a_2/a_1)V^B] \Mod{\Lambda_E^A({\bf a}, {\boldsymbol\beta})} \nonumber \\
&=&[T^A+(a_2/a_1)V^B-(a_2/a_1)V^B] \Mod{\Lambda_E^A({\bf a}, {\boldsymbol\beta})}  \nonumber \\
&= & [T^A] \Mod{\Lambda_E^A({\bf a}, {\boldsymbol\beta})} \nonumber \\
&=& T^A.
\end{eqnarray}
Since $T^A$ is reliably decoded, $W^A$ can then be retrieved.

\end{itemize}

\subsubsection{Information Leakage Rate}
Here, we proof that the LC scheme is information theoretically secure.

\lemma{For any $\delta>0$, if 
\begin{equation}\label{eq:conrsa}
R_t^B({\bf a}, {\boldsymbol\beta})=\hat{R}_t^B({\bf a}, {\boldsymbol\beta}),
\end{equation}
 there exist a sequence of $({\bf a}, {\boldsymbol\beta})$ SCF lattice chain codes which achieves
\begin{equation}
\frac{1}{N}I(W^A;Y^R)<\delta.
\end{equation}}\label{lm:2}

The proof of this lemma is in Appendix~\ref{ap:2}.

\subsubsection{Achievable Secrecy Rate}

Similar as the previous section, we also distinct three cases w.r.t. $P_R$ and $\sigma$. For each case, we specify the settings of ${\bf a}$, ${\boldsymbol\beta}$, and $R_t^i({\bf a}, {\boldsymbol\beta})$.
\begin{itemize}

\item $P_R \geq P_A$ and $\sigma \leq \underline{\sigma}$.

In this case, similar to the RB scheme, we can set $R_t^i({\bf a}, {\boldsymbol\beta})$ equals to $\hat{R}_t^i({\bf a}, {\boldsymbol\beta})$ and set $\hat{R}_t^i({\bf a}, {\boldsymbol\beta})$ accordingly to (\ref{eq:rtcf}). Combining with (\ref{eq:il4}), (\ref{eq:macab}), and (\ref{eq:rta}), we achieve any secrecy rate satisfying (\ref{eq:rbrs1}) by choosing sufficiently large $N$ and sufficiently small $\delta'$.

\item $P_R \geq P_A$ and $\sigma > \underline{\sigma}$.

In this case, if we use the same settings as the previous case, (\ref{eq:1c}) will be violated. Moreover, unlike the RB scheme, due to the constraint of the $({\bf a}, {\boldsymbol\beta})$ SCF lattice chain code, using the same lattice codebook with a simple decreasing of the transmit rate will violate (\ref{eq:conrsa}). Hence a new lattice codebook with different ${\bf a}, {\boldsymbol\beta}$ should be generated w.r.t. the constraint
\begin{equation}\label{eq:conrcfb}
R_{\rm CF}^B({\bf a}, {\boldsymbol\beta}) \leq C(P_B/\sigma^2)
\end{equation}
and $\hat{R}_t^i({\bf a}, {\boldsymbol\beta})$ should be set accordingly to (\ref{eq:rtcf}). Then, we set $R_t^i({\bf a}, {\boldsymbol\beta})$ equal to $\hat{R}_t^i({\bf a}, {\boldsymbol\beta})$.
Thus, any secrecy rate satisfying
\begin{equation}\label{eq:lcrs}
R_s < \max_{{\bf a},{\boldsymbol\beta}:R_{\rm CF}^B({\bf a}, {\boldsymbol\beta}) \leq C(P_B/\sigma^2)}R_s({\bf a}, {\boldsymbol\beta})
\end{equation}
is achievable.

\item $P_R < P_A$.

In this case, unlike the RB scheme, the LC scheme guarantees a reliable transmission of $T^A$ as long as (\ref{eq:hatpa}) holds. Note that our scheme holds for any pair of ${\bf a}, {\boldsymbol\beta}$. Moreover, as long as $P_R < P_A$, by \cite{jingge}, for any secrecy rate $\hat{R}_s^A({\bf a}, {\boldsymbol\beta})$ satisfying (\ref{eq:hatpa}), there exists a pair of ${\bf a}, {\boldsymbol\beta}$ which achieves that rate. Hence, by choosing ${\bf a}, {\boldsymbol\beta}$, and the codebook accordingly to (\ref{eq:hatpa}), we can straightforwardly achieve any secrecy rate smaller than $C(P_R)$.

\end{itemize}

Combining the three cases we have the following lemma.
\theorem{For a two-hop channel with an untrusted relay, with the LC scheme, any secrecy rate $R_s$ satisfying
\begin{equation}\label{eq:lcrs3}
R_s < \min (\max_{{\bf a}, {\boldsymbol\beta}}R_s({\bf a}, {\boldsymbol\beta}),C(P_R))
\end{equation}
is achievable if $\sigma \leq \underline{\sigma}$ and
\begin{equation}\label{eq:lcrs4}
R_s < \min ( \max_{{\bf a},{\boldsymbol\beta}:R_{\rm CF}^B({\bf a}, {\boldsymbol\beta}) \leq C(P_B/\sigma^2)}R_s({\bf a}, {\boldsymbol\beta}), C(P_R) )
\end{equation}
is achievable if $\sigma > \underline{\sigma}$.}\label{lm:lc}

\subsection{Achievable Secrecy Rates for Special Channel Configurations}

Here, we consider two special channel configurations. Firstly, if $P_R \geq P_A$ and the jammer is collocated with the destination, i.e., $\sigma=0$, we can maximize the first term in (\ref{eq:scr1}) by choosing ${\bf a}=(1,1)$ and ${\boldsymbol\beta}=(\sqrt{\frac{P_B(1+P_A)}{P_A(1+P_B)}},1)$, which gives us the following corollary.

\corollary{On a two-hop channel with an untrusted relay, if $P_R \geq P_A$ and $\sigma=0$, any secrecy rate $R_s$ satisfying 
\begin{equation}
R_s < \frac{1}{2}\log_2\frac{1+P_A+P_B}{(\sqrt{(1+P_A)(1+P_B)}-\sqrt{P_AP_B})^2}-1
\end{equation}
is achievable.}\label{co:1}

Then, if $P_A=P_B=P_R$ and $\sigma=0$, by Corollary~\ref{co:1} we straightforwardly have the following corollary.

\corollary{On a two-hop channel with an untrusted relay, if $\sigma=0$ and $P_A=P_B=P_R$, any secrecy rate $R_s$ satisfying
\begin{equation}\label{eq:co2}
R_s < \frac{1}{2}\log_2(\frac{1}{2}+P_A)-\frac{1}{2}
\end{equation}
is achievable.}\label{co:2}


\subsection{Comparison Between the Two Schemes}

Firstly, we compare the two schemes in terms of simplicity in deployment, the LC scheme surely enjoys the benefit of a simpler structure and decoding. Also, the RB scheme requires a very long sequence of lattice codewords to achieve secrecy, i.e. $N$ and length $l$ should be sufficiently large, which is not the case for the LC scheme in which only $N$ needs to be chosen sufficiently large.

\begin{figure*}[!ht]
  \subfigure[The comparison under the limited relay power condition (destination functions as the jammer).]{ 
    \label{fig:sac1} 
     \includegraphics[width=0.45\textwidth]{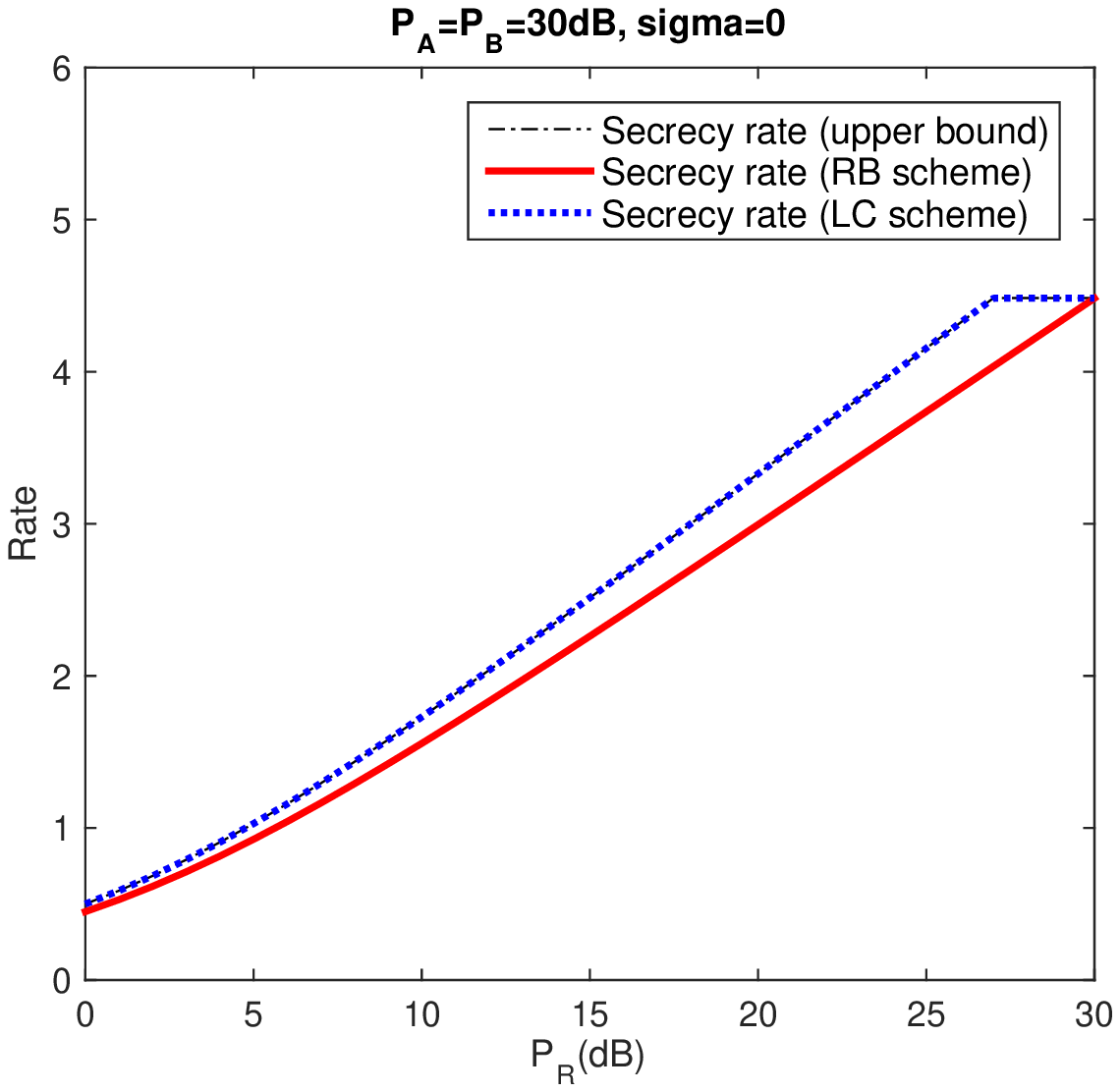}}
  \subfigure[The comparison under the non-collocated jammer and destination condition.]{ 
    \label{fig:sac2} 
     \includegraphics[width=0.45\textwidth]{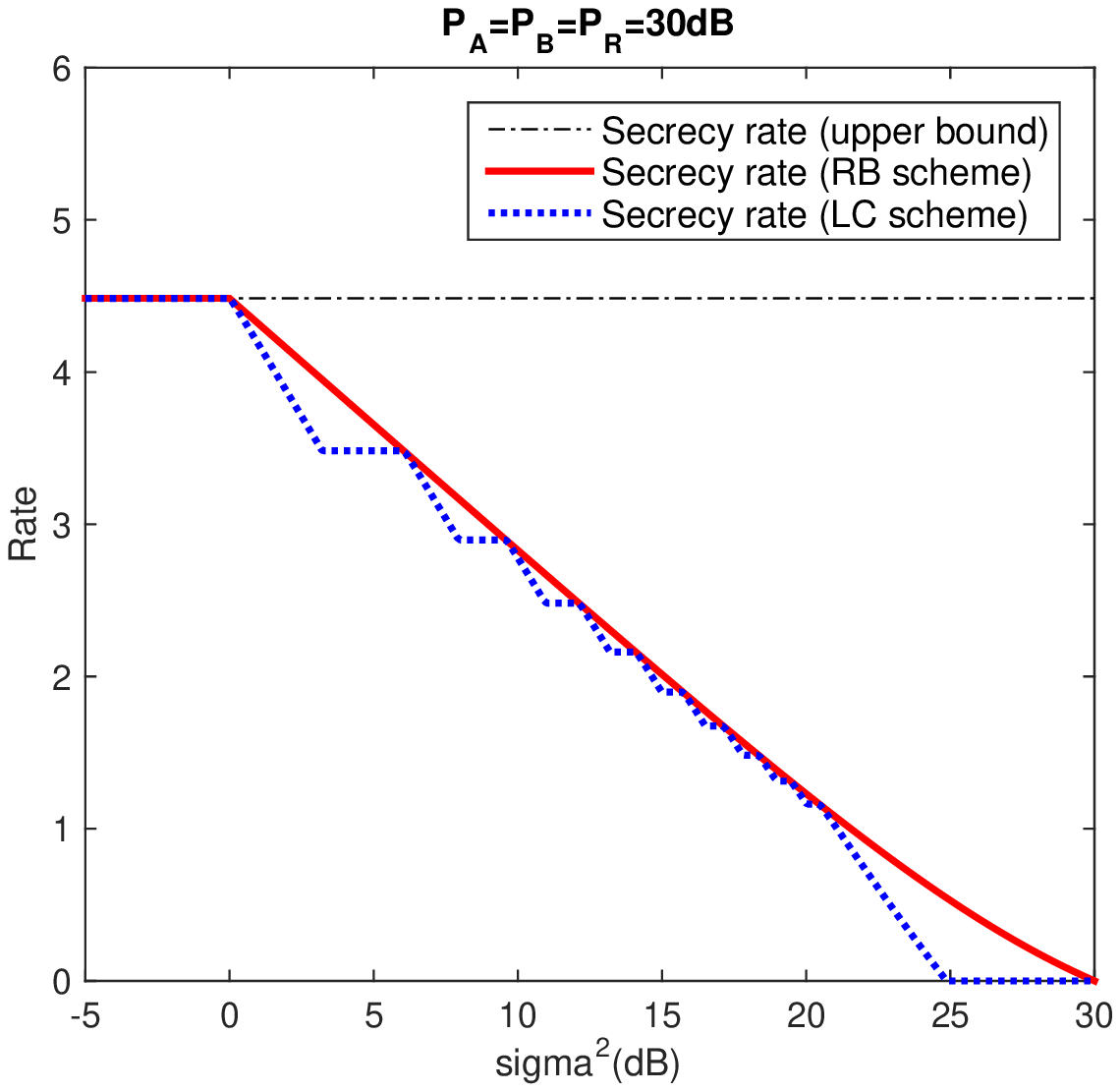}}

  \subfigure[The comparison under both conditions]{ 
    \label{fig:sac3} 
     \includegraphics[width=0.45\textwidth]{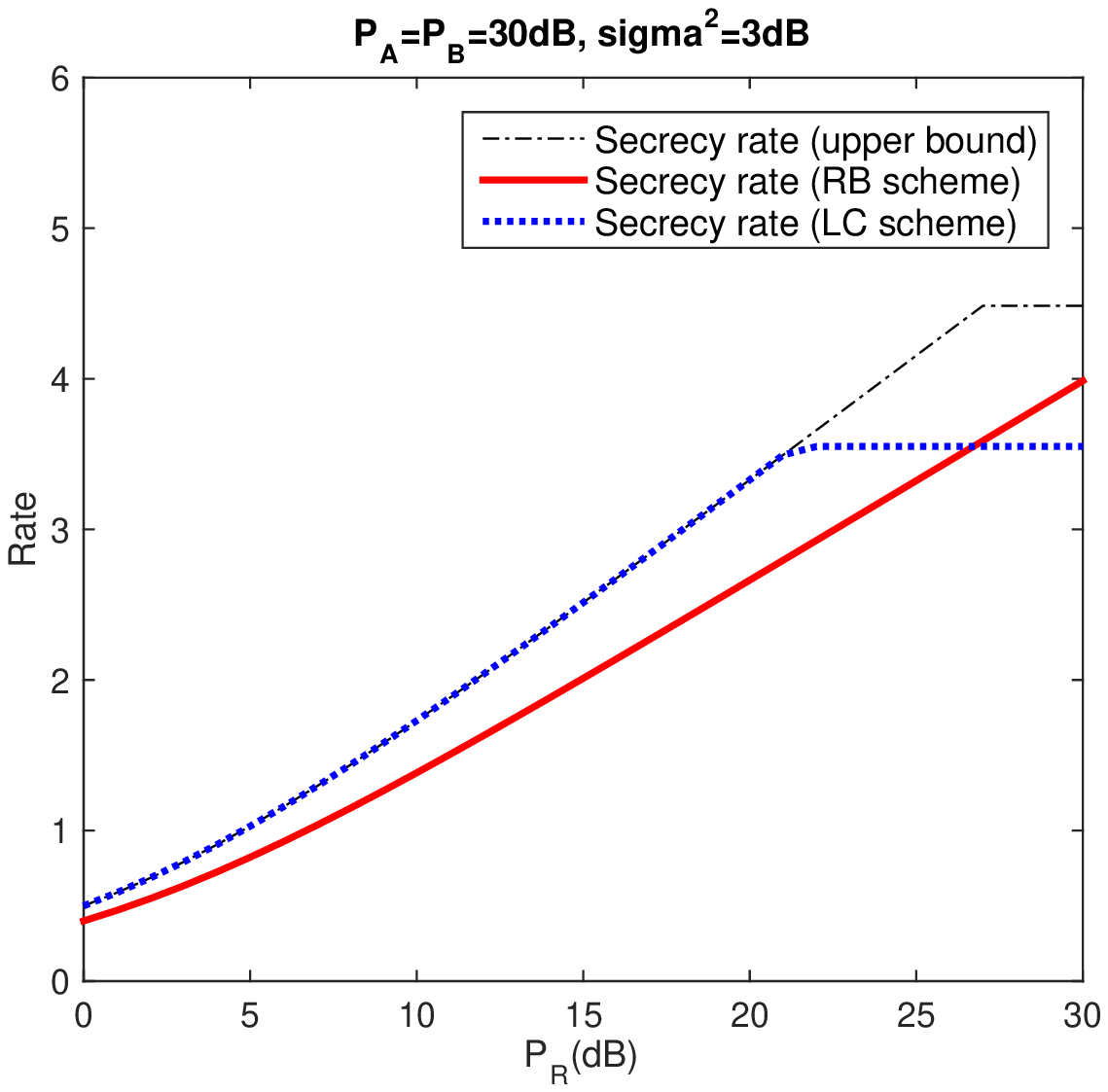}}
\caption{Comparison between the achievable secrecy rate of the RB scheme and the LC Scheme}
\end{figure*}

Then, we compare the achievable secrecy rates of the two schemes. In the case of $P_R \geq P_A$ and $\sigma \leq \underline{\sigma}$, both of the two schemes achieve the same secrecy rate of (\ref{eq:rbrs1}). Then, when $P_R < P_A$, thanks to the chain structure, the LC scheme allows the relay to save the part of the energy of transmitting the random vector $V^A$. This feature allows the LC scheme to achieve a secrecy rate that equals the capacity when the relay has limited power, while the RB scheme underperforms. In other word, when $\sigma \leq \underline{\sigma}$, the rate of the LC scheme (\ref{eq:lcrs3}) is always no lower than the rate of the RB scheme (\ref{eq:rsrb11}). In Fig.~\ref{fig:sac1} where $P_A=P_B=30{\rm dB}$ and the destination is used as the jammer, it is clear that the curve of the LC scheme is higher than the RB scheme and coincides with the upper bound. Note that in this paper all the powers are shown in dB, since they are actually the SNR with unit noise.

In the case of $P_R \geq P_A$ and $\sigma > \underline{\sigma}$, for the sake of reliable decoding of $V^B$ at node $C$, the transmit rate should be reduced. For the LC scheme, due to the constraint of (\ref{eq:conrsa}), node $B$ cannot simply use the same lattice codebook and reduce its transmit rate. Hence, in this case, the RB scheme could achieve a higher rate than the LC scheme. In Fig.~\ref{fig:sac2}, the rate of the RB scheme is always higher than the LC scheme. However, they are both far away from the upper bound when $\sigma$ is large. The shape of the curve of the LC scheme is due to the fact that sometimes the achievable secrecy rate is maximized by choosing a different $a_1$ for the different $\sigma$, which is a positive integer.

In the case of $P_R < P_A$ and $\sigma> \underline{\sigma}$, it will be a trade-off between these two issues. As observed in Fig.~\ref{fig:sac3}, when $\sigma^2=3$dB, the LC scheme performs better when $P_R$ is low, but is overtaken by the RB scheme when $P_R$ is larger than some threshold.


Summarizing the three cases discussed above, a new lower bound on the achievable secrecy rate on this channel is derived.
\corollary{On a two-hop channel with an untrusted relay, any secrecy rate $R_s$ satisfying
\begin{equation}\label{eq:scr1}
R_s < \min\left(\max_{{\bf a}, {\boldsymbol\beta}}R_s({\bf a}, {\boldsymbol\beta}), C(P_R)\right)
\end{equation}
is achievable if $\sigma \leq \underline{\sigma}$, and 
\begin{eqnarray}
R_s &< &\max \left[ \min\left( \max_{{\bf a}, {\boldsymbol\beta}:R_{\rm CF}^B({\bf a}, {\boldsymbol\beta}) \leq C(P_B/\sigma^2)}R_s({\bf a}, {\boldsymbol\beta}), C(P_R)\right) \right., \nonumber \\
&&\hspace{2cm} \left. \left(C(P_A)+ C(P_B/\sigma^2)-C(P_A+P_B)\right)\frac{C(P_R)}{C(P_A)}\right]  \label{eq:scr2}
\end{eqnarray}
is achievable if $\sigma > \underline{\sigma}$.}\label{th:1}

\section{Performance Analysis and Comparison}\label{s:comp}

In this section we compare the achievable secrecy rate of our schemes and other schemes under various scenarios.

\subsection{Symmetric Two-hop Channel with Destination as Jammer}

We first discuss the very well studied symmetric two-hop channel with the destination functioning as a cooperative jammer, which is a special case of our model when $P_A=P_R=P_B$ and $\sigma^2=0$. Here, both our schemes achieve the sames secrecy rate of (\ref{eq:co2}). We compare it to the achievable secrecy rate with an amplify-and-forward based scheme proposed by Sun {\em et al.} in \cite{sun} and a modulo-and-forward based scheme propose by Zhang {\em et al.} in \cite{zhang}. Their achievable secrecy rates are in (\ref{eq:sun}) and (\ref{eq:zhang}), respectively. In particular, (\ref{eq:zhang}) can be simplified to
\begin{equation}
R_s < C(P_A)-\frac{1}{2}-\frac{1}{2}\log_2\left(1+\frac{P_A}{1+P_A}\right).
\end{equation}

We also compare our schemes with the compress-and-forward based scheme proposed by He {\em et al.} in \cite{he}.The achievable secrecy rate is in (\ref{eq:he1}) and can be simplified to
\begin{equation}
R_s < \frac{1}{2}\log_2\left(2+\frac{1}{P_A}+P_A\right)-1.
\end{equation}

\begin{figure}[H]
\centering
\includegraphics[width=0.5\textwidth]{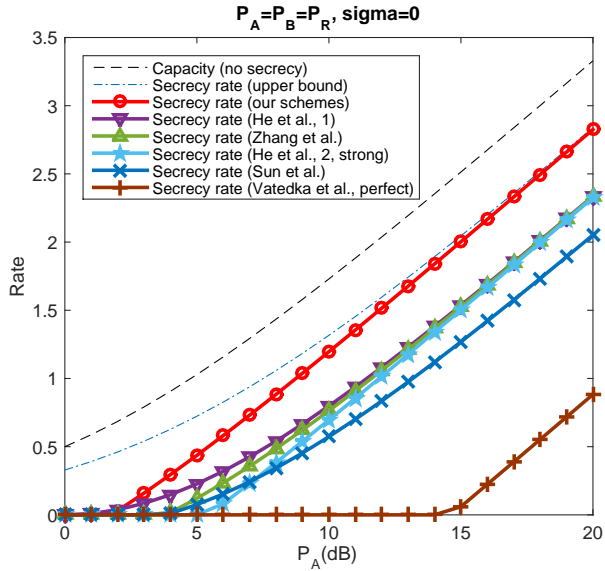}
\caption{Comparison between the achievable secrecy rates of variant schemes in a symmetric two-hop channel using the destination as jammer.}
\label{fig:symcom}
\end{figure}

In Fig.~\ref{fig:symcom}, we set $P_A=P_B=P_R=20$dB and compare these schemes with the upper bound (\ref{eq:ubsec}). Moreover, we show the rate of He {\em et al.} in \cite{he3} and Vatedka {\em et al.} in \cite{vatedka} in the same figure, although these are the rates for strong secrecy and perfect secrecy, respectively. We also show the capacity without the consideration of secrecy as a reference. It is clear that our scheme outperforms all other existing secure transmission scheme in the high SNR region and is upper bound achieving when $P_A\to \infty$. Also, it is interesting to observe that, in the high SNR region, to achieve strong secrecy and perfect secrecy, a rate of $0.5$ and $0.5+\log_2e$ bits/channel use is lost, respectively.

\subsection{Asymmetric Two-hop Channel with Destination as Jammer}

\begin{figure*}[!ht]
\centering
  \subfigure[$P_A\leq P_B=P_R=20$dB.]{ 
    \label{fig:asymcom1} 
     \includegraphics[width=0.45\textwidth]{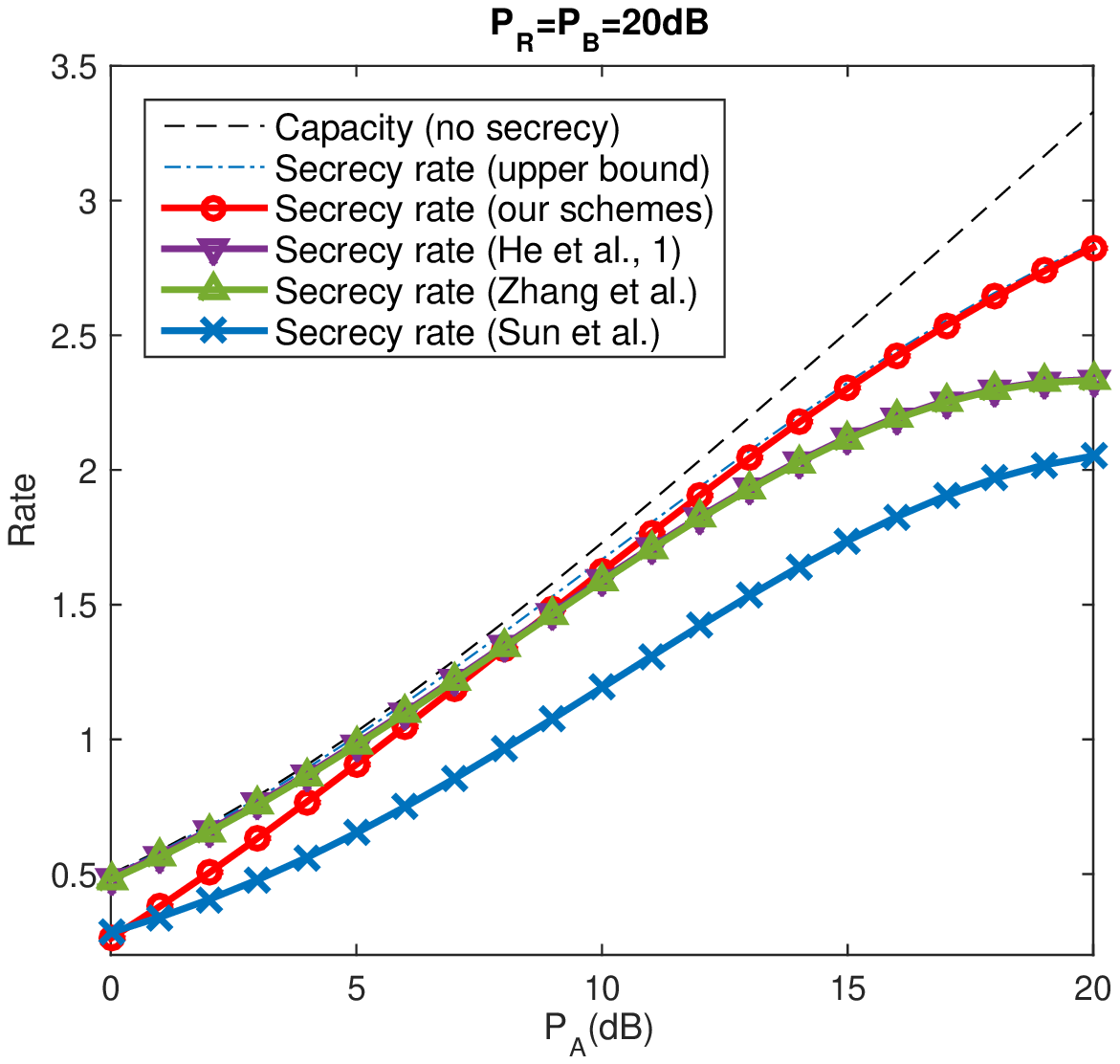}}
  \subfigure[$P_B\leq P_A=P_R=20$dB.]{ 
    \label{fig:asymcom2} 
     \includegraphics[width=0.45\textwidth]{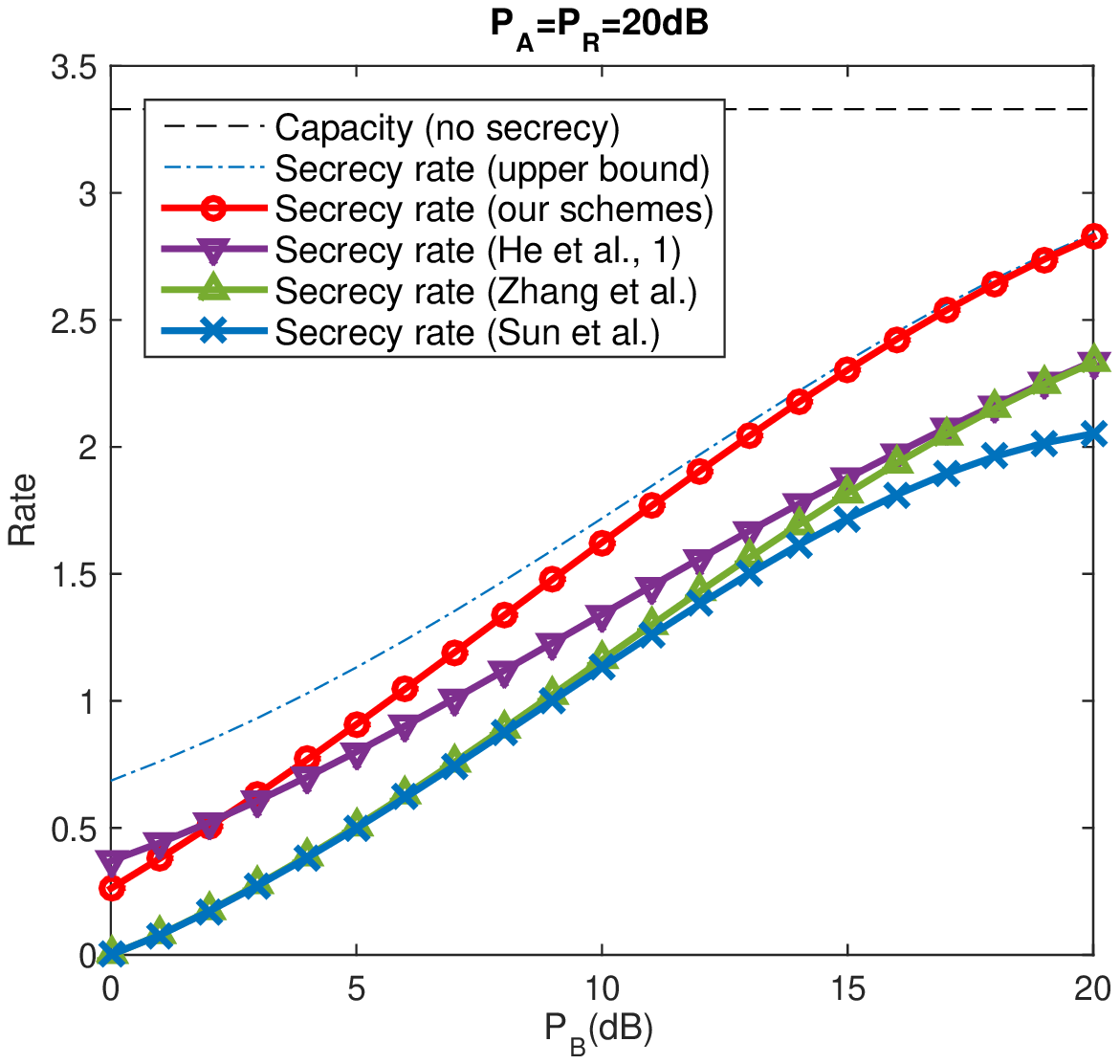}}
  
\subfigure[$P_R\leq P_A=P_B=20$dB.]{
    \label{fig:asymcom3} 
     \includegraphics[width=0.45\textwidth]{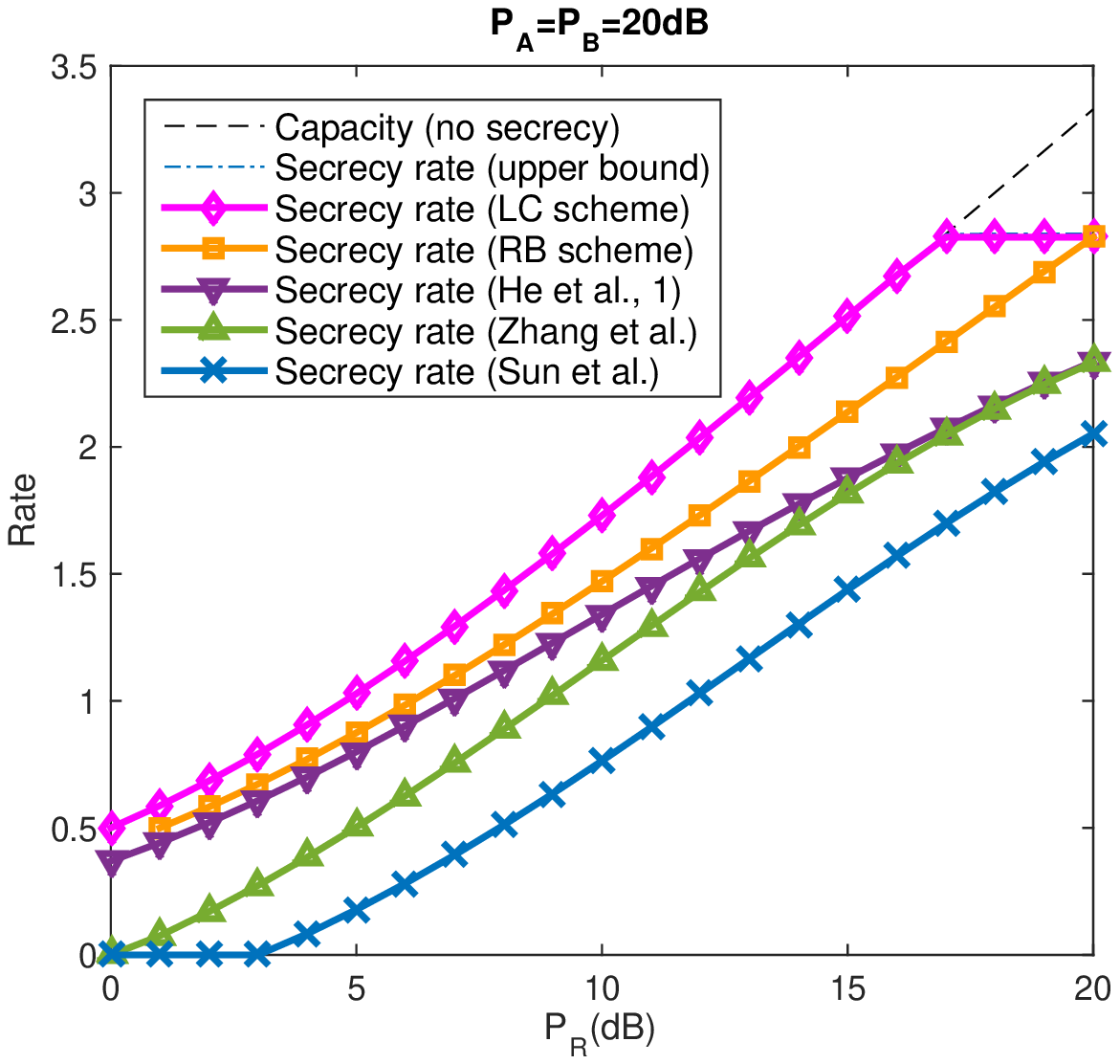}}
\subfigure[$P_B=10P_A,P_R=P_A$.]{ 
    \label{fig:asymcom4} 
     \includegraphics[width=0.45\textwidth]{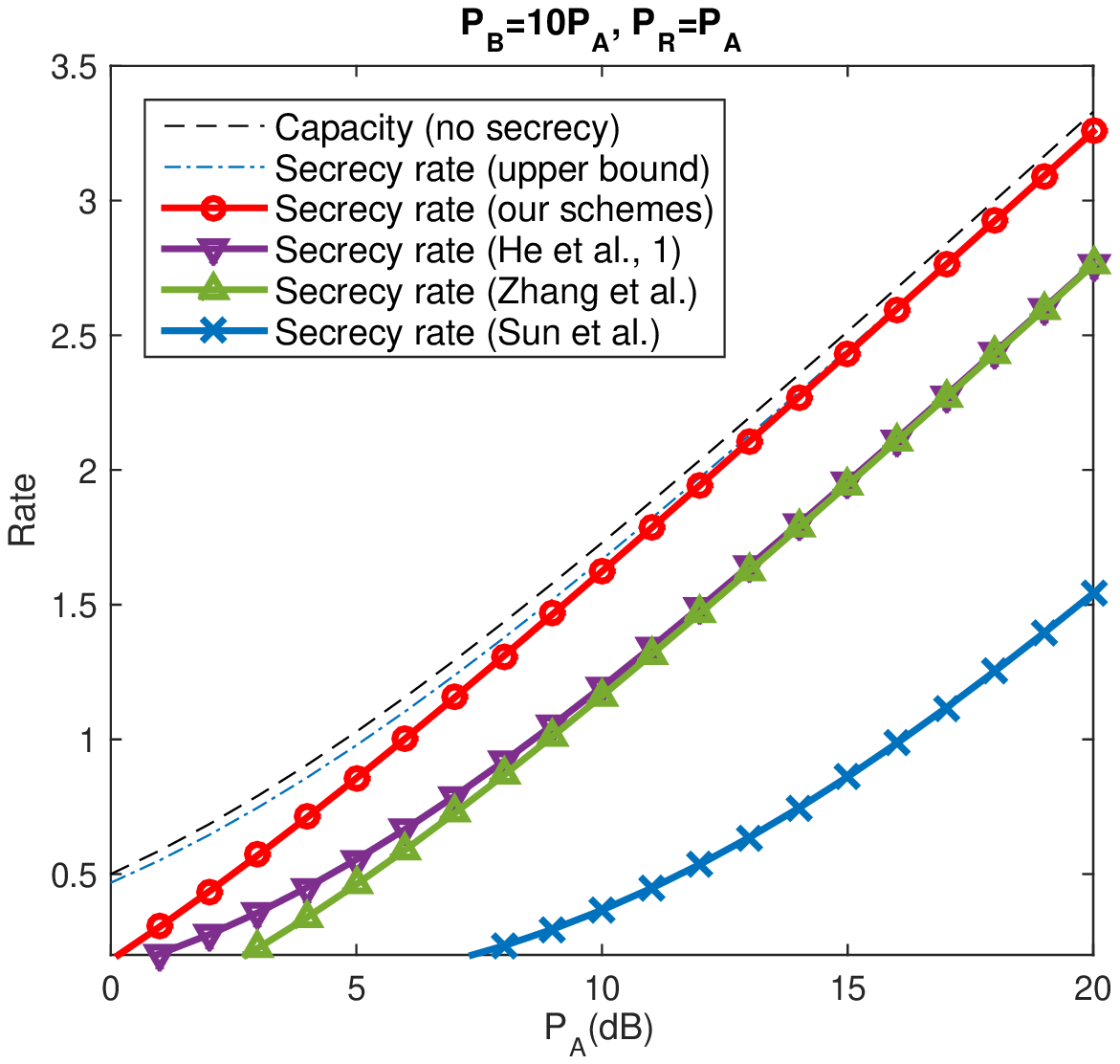}}
\caption{Comparison between the achievable secrecy rates of various schemes in asymmetric two-hop channels using the destination as jammer.}
\end{figure*}
\begin{figure*}[!ht]
\centering
\subfigure[$P_B=0.1P_A,P_R=P_A.$]{ 
    \label{fig:asymcom5} 
     \includegraphics[width=0.45\textwidth]{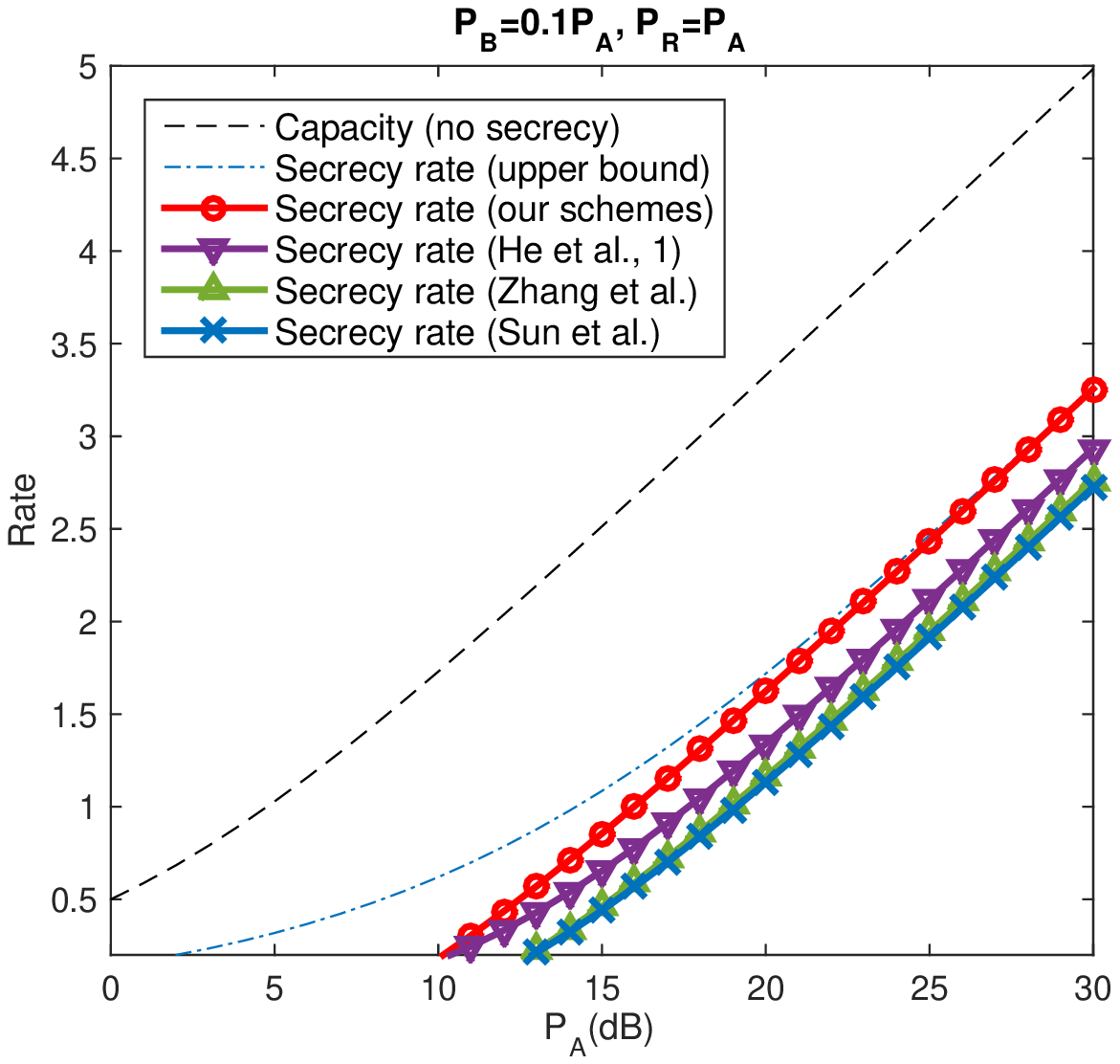}}
\subfigure[$P_B=0.1P_A, P_R=20$dB.]{ 
    \label{fig:asymcom6} 
     \includegraphics[width=0.45\textwidth]{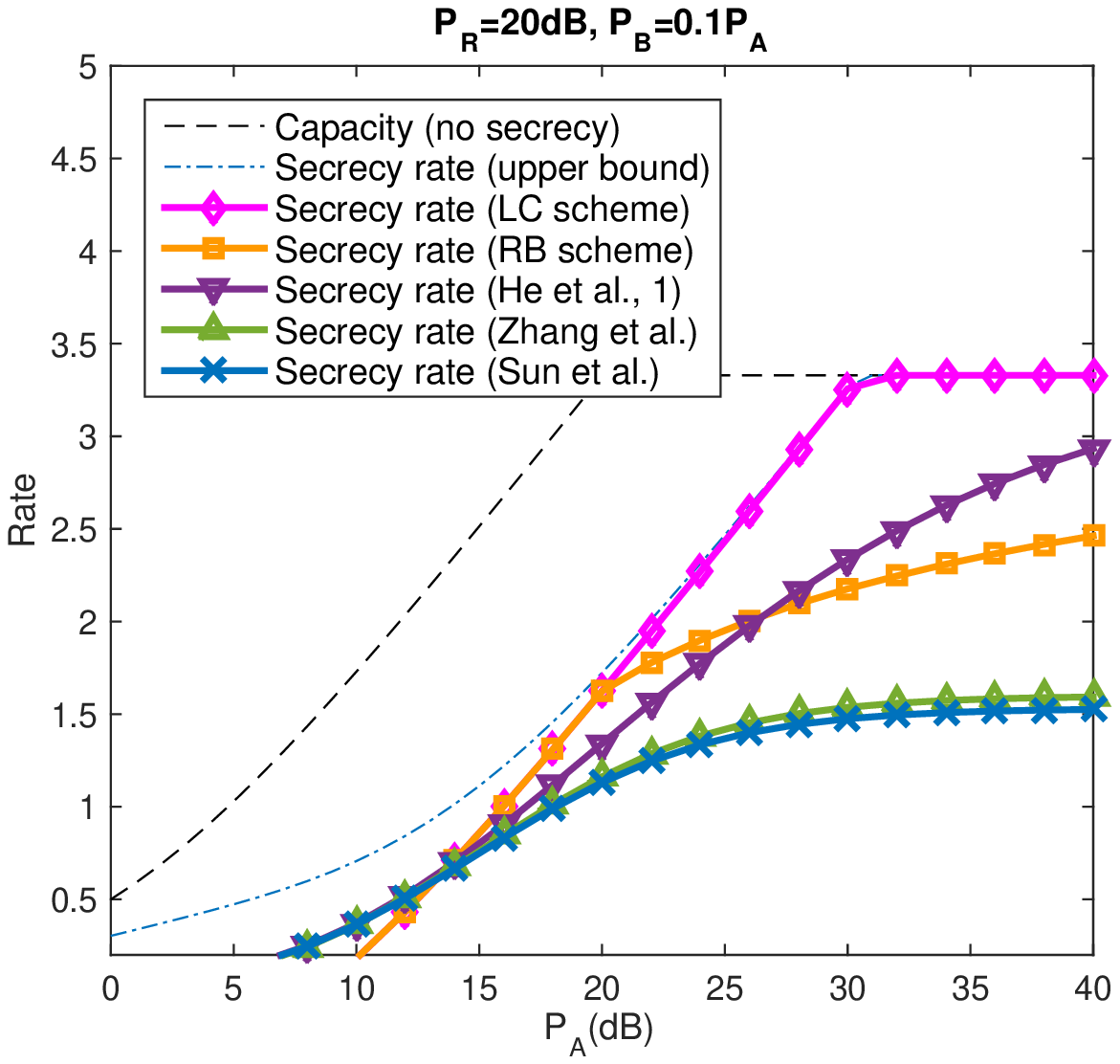}}
\caption{Comparison between the achievable secrecy rates of various schemes in asymmetric two-hop channels using destination as jammer (Continued).}
\end{figure*}
In case of $\sigma=0$, we compare our schemes to the capacity without secrecy constraint, the upper bound (\ref{eq:ubsec}), as well as the schemes by Sun {\em et al.} in \cite{sun}, Zhang {\em et al.} in \cite{zhang}, and He {\em et al.} in \cite{he}, the rates of which are in (\ref{eq:sun}), (\ref{eq:zhang}), and (\ref{eq:he1}), respectively. It can be observed from Fig~\ref{fig:asymcom1}-\ref{fig:asymcom3} that if we fix two of $P_i, i\in \{A,B,R\}$ and change one of them, our schemes outperform all other schemes except for the low source and/or jammer power case. Moreover, the LC scheme achieves the upper bound for the low relay power cases (the curve coincides with the upper bound). To the best of our knowledge, this is the first upper bound achieving scheme for the limited relay power and non-infinity source power case. 

Furthermore, we compare the achievable secrecy rate of various schemes with the upper bound in the case of $P_B=\alpha P_A, \alpha >0$, and $P_A\to \infty$. In this case, we define the gap between the upper bound of the secrecy rate derived in \cite{he} and the channel capacity without secrecy consideration as
\begin{eqnarray}
G_0 &=&\lim_{P_A \to \infty} [\min (C(P_A),C(P_R))-\min (R_b,C(P_R))],
\end{eqnarray}
where $R_b$ is the upper bound given in (\ref{eq:ubsec}). Note that this upper bound is only for the secrecy rate in Phase 1. In Phase 2, the secrecy rate is upper bounded by $C(P_R)$.
Similarly, for each secure transmission scheme, we define the gap between the achievable secrecy rate and the capacity without secrecy consideration as
\begin{eqnarray}
G &=&\lim_{P_A \to \infty} [\min (C(P_A),C(P_R)) -\limsup_{N\to\infty} R_s],
\end{eqnarray}
where $R_s$ is the achievable secrecy rate of the scheme.

When $P_R \geq P_A$, both our schemes achieve any rate satisfying (\ref{eq:rbrs1}), in which the RHS equals the RHS of (\ref{eq:rsmax}). It can be calculated that we have $G=G_0=C(1/\alpha)$, which reflects that our schemes are upper bound achieving in this case. Then, when $P_R < P_A$, the LC scheme still achieves the upper bound, which in this case is the channel capacity without secrecy consideration, i.e., $G=G_0=0$.

The $G_0$ value as well as the $G$ values of various secure transmission schemes are shown in Table~\ref{tb:app} for some channel configurations. Here, $\gamma$ is defined as a positive real number. Is is shown that the LC scheme is the only upper bound achieving scheme in all the three cases considered in the table. For all other existing schemes, there are always gaps of at least a constant between the achievable secrecy rate and the upper bound in one or more cases.

\begin{table}[!ht]
\centering
\begin{threeparttable}
\begin{tabular}{|l|l|l|l|@{}m{0pt}@{}}
\hline \hline
& \multicolumn{2}{c|}{$P_R = \gamma P_A$}& \multicolumn{1}{c|}{$P_R$ fixed} \\ \cline{2-3}
& $\gamma<1$ & $\gamma \geq 1$ & \\ \hline
$G_0$ (Upper bound \cite{he}) & 0 &$C(\frac{1}{\alpha})$ & 0 & \\[5pt] \hline
$G$ (RB scheme) & $\frac{C(P_R)}{C(P_A)}C(\frac{1}{\alpha})$ &$C(\frac{1}{\alpha})$ & 0 & \\[5pt] \hline
$G$ (LC scheme) & 0 &$C(\frac{1}{\alpha})$ & 0 & \\[5pt] \hline
$G$ (He's scheme \cite{he})& $C(\frac{1}{\alpha})+C(\gamma)$ & $[C(\frac{1}{\alpha}), C(\frac{1}{\alpha})+C(\frac{1}{\gamma})] $ \tnote{a} & 0 &\\[5pt] \hline
$G$ (Zhang's scheme \cite{zhang})& $C(\frac{1}{\alpha})+C(\gamma)$ &$C(\frac{1}{\alpha} )+C(\frac{1}{\gamma})$  & $C(\frac{1}{\alpha})$  &\\[5pt] \hline
$G$ (Sun's scheme \cite{sun})& $C(\frac{1}{\alpha})+C(\gamma+\alpha)$ &$C(\frac{1}{\alpha} )+C(\frac{\alpha+1}{\gamma})$ & $C(\frac{1}{\alpha} )+C(\frac{\alpha P_R}{P_R+\alpha+1})$  &\\[5pt] \hline \hline
\end{tabular}
\begin{tablenotes}
\item[a] {\small If $\alpha\geq 1$, the $G$ value is $C(\frac{1}{\alpha} )+C(\frac{1}{\gamma})$. However, if $\alpha <1$, the $G$ value can be smaller than $C(\frac{1}{\alpha} )+C(\frac{1}{\gamma})$ depending on $\alpha$. Hence, we only show the interval for the rate here.}
\end{tablenotes}
\end{threeparttable}
\caption{The $G_0$ and $G$ values for various schemes when $P_B=\alpha P_A$ and $P_A \to \infty$.}\label{tb:app}
\end{table}

In Fig.~\ref{fig:asymcom4} and \ref{fig:asymcom5} we show the cases of $\gamma=1, \alpha=10$ and $\gamma=1, \alpha=0.1$, respectively. In Fig.~\ref{fig:asymcom6}, we fix $P_R=20{\rm dB}, P_B=0.1P_A$ and show the performance of various schemes when $P_R$ is limited.



\subsection{External Jammer}

\begin{figure*}[!ht]
\centering
\subfigure[$\sigma\geq -5 {\rm dB}, P_A=P_B=P_R=30$dB.]{ 
    \label{fig:exjcom1} 
     \includegraphics[width=0.45\textwidth]{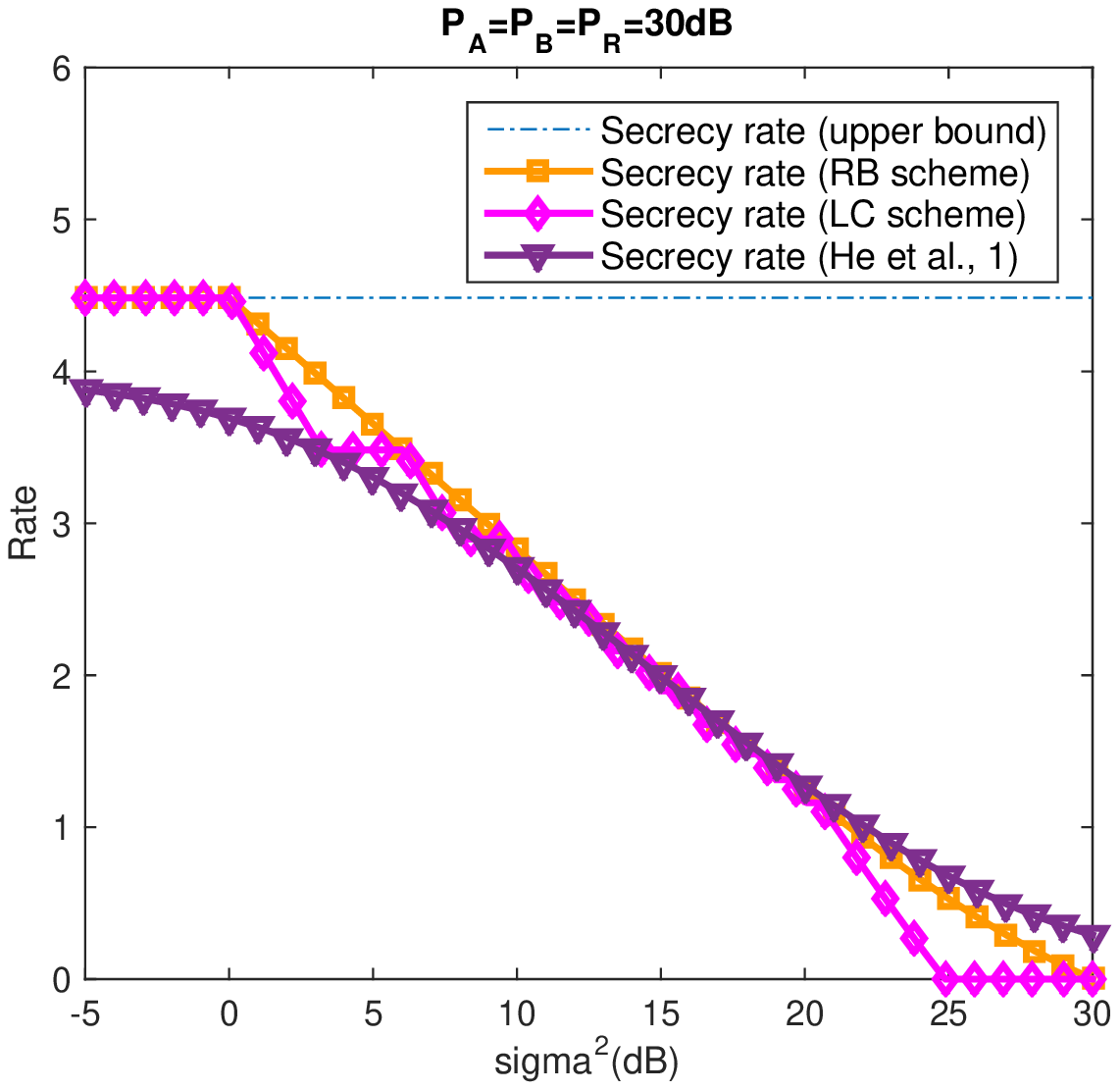}}
\subfigure[$\sigma^2=3{\rm dB}$ and $P_R\leq P_A=P_B=20$dB.]{ 
    \label{fig:exjcom2} 
     \includegraphics[width=0.45\textwidth]{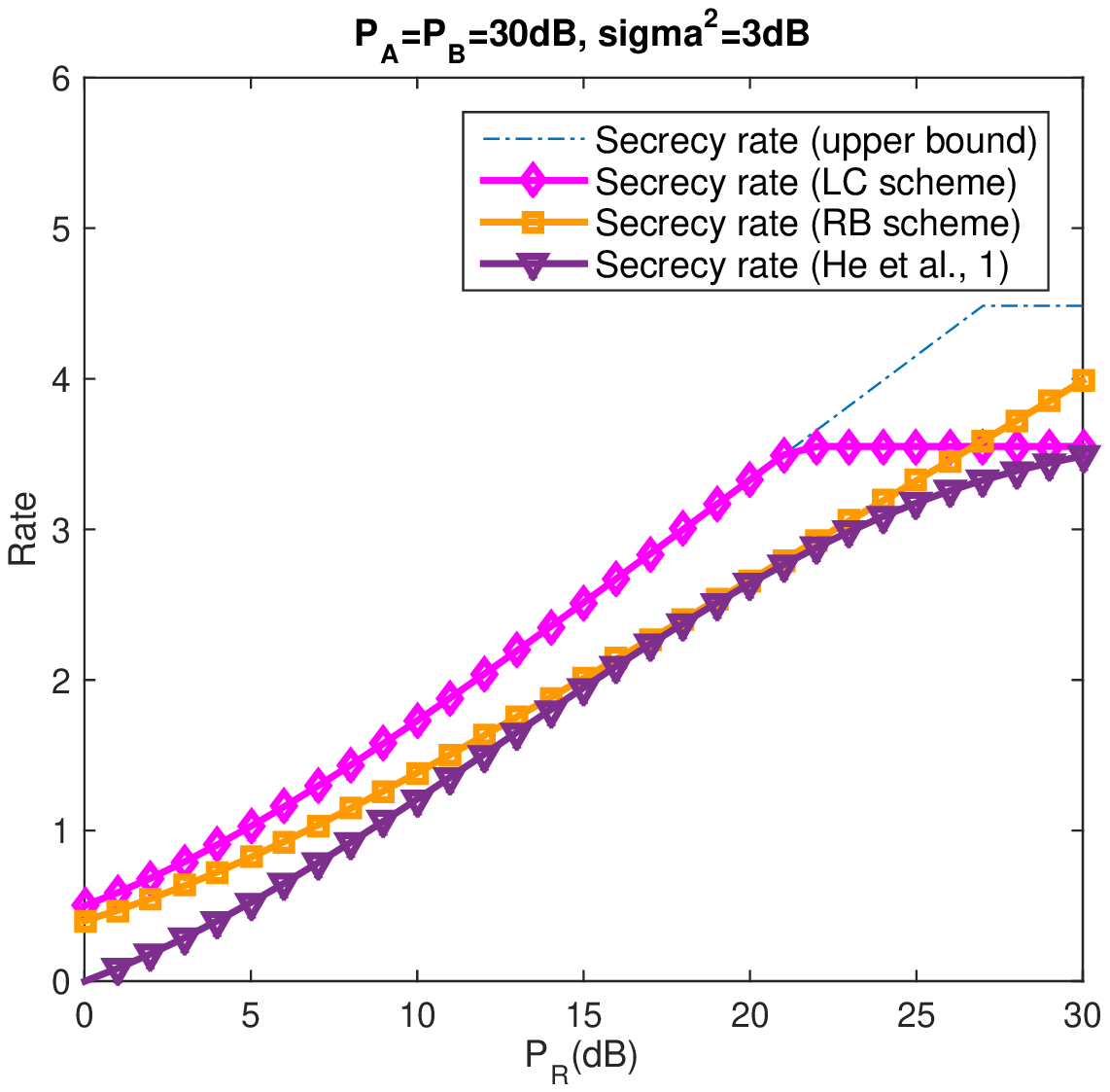}}
\caption{Comparison between the achievable secrecy rates of various schemes in two-hop channels with an external jammer.}
\end{figure*}

Since the external jammer case is only considered by He {\em et al.} in \cite{he}, we compare our schemes to their scheme in Fig.~\ref{fig:exjcom1} and Fig.~\ref{fig:exjcom2} for different $\sigma^2$ and $P_R$. In Fig.~\ref{fig:exjcom1} it can be observed that our schemes perform better when $\sigma$ is small. When the channel between $B$ and $C$ is too noisy, the scheme of He {\em et al.} achieves a better rate. In Fig.~\ref{fig:exjcom2}, it is shown that our schemes have better performance in the limited relay power case. In particular, if the relay power is very low, the LC scheme is upper bound achieving even for large $\sigma$.

\section{Two-Hop Channel with an Eavesdropper}\label{s:ee}
In this section, we propose another channel model, in which we consider the case that the relay is honest and cooperative, but there is an external eavesdropper. 

\subsection{Model}

We firstly consider a two-hop channel in which node $A$ wants to transmit information to node $B$ using a relay node $R$ to forward the information. During the process an eavesdropper is trying to obtain the information transmitted by node $A$. In this model we assume that the destination $B$ also functions as a cooperative jammer. We also assume that the communication takes places over two phases, each including $N$ channel uses. We use $X^A, X^B, X^R \in \mathbb{R}^N$ for the transmissions of node $A$, the destination $B$, and the relay $R$, respectively. We use $Y^R, Y^E_1, Y^E_2, Y^B\in \mathbb{R}^N$ for the receptions of the relay, the eavesdropper in the two phases, and node $B$, respectively. In the first phase, node $A$ transmits to the relay $R$ and this transmission is eavesdropped by the eavesdropper $E$. The destination $B$ simultaneously transmits a jamming signal to confuse the eavesdropper, which is also superimposed with the transmission of node $A$ at the relay $R$. Hence, we have
\begin{eqnarray}
Y^R&=&h_1X^A+h_2X^B+Z^R, \label{eq:yrx}\\
Y^E_1&=&h_1'X^A+h_2'X^B+Z^E_1,
\end{eqnarray}
where $Z^R$ and $Z^E_1$ are $N$-dimensional independent Gaussian noise vectors and $h_1,h_2,h_1',h_2' \in \mathbb{R}^+$ are the channel coefficients. We further denote ${\bf h}=(h_1,h_2)$ and ${\bf h}'=(h_1', h_2')$.

\begin{figure}[!ht]
\centering
\tikzstyle{vertex}=[circle,minimum size=30pt, inner sep=0pt, draw]
\tikzstyle{edge}=[-latex, black, thick]
\tikzstyle{dot}=[edge, dotted]
\tikzstyle{dash}=[edge, densely dashed]
\tikzstyle{loosedo}=[edge, loosely dotted]
\subfigure[Phase 1]{ 
\begin{tikzpicture}[scale=2.8]
\node[vertex] (t1) at (0,2.4)  {$A$} ;
\node[vertex] (r) at (1,2.4) {$R$};
\node[vertex] (t2) at (2,2.4) {$B$} ;
\node[vertex] (j) at (1,3) {$E$};

\draw[edge, above] (0.2,2.7) to node {$h_1$} (0.8, 2.7);
\draw[dot, above](0.2,2.7) to node {$h_1'$} (0.8, 3.2);
\draw[dot, above](1.8,2.7) to node {$h_2'$} (1.2, 3.2);
\draw[edge, above](1.8, 2.7) to node {$h_2$} (1.2, 2.7);


\node (xa) at (0, 2.7) {$X^A$};
\node (yr) at (1, 2.7) {$Y^R$};
\node (xb) at (1, 3.3) {$Y^E_1$};
\node (yc1) at (2, 2.7) {$X^B$};
\end{tikzpicture}}

\subfigure[Phase 2]{ 
\begin{tikzpicture}[scale=2.8]
\node[vertex] (t1) at (0,2.4)  {$A$} ;
\node[vertex] (r) at (1,2.4) {$R$};
\node[vertex] (t2) at (2,2.4) {$B$} ;
\node[vertex] (j) at (1,3) {$E$};

\draw[edge, right](1, 2.6) to node {$h_3$} (1, 2.8);
\draw[edge, above](1.2, 2.5) to node {$h_2$} (1.8, 2.5);


\node (yr) at (0.8, 2.7) {$X^R$};
\node (xb) at (1, 3.3) {$Y^E_2$};
\node (yc1) at (2, 2.7) {$Y^B$};
\end{tikzpicture}}
\caption{Two-hop channel with an eavesdropper.}\label{fig:exte}
\end{figure}
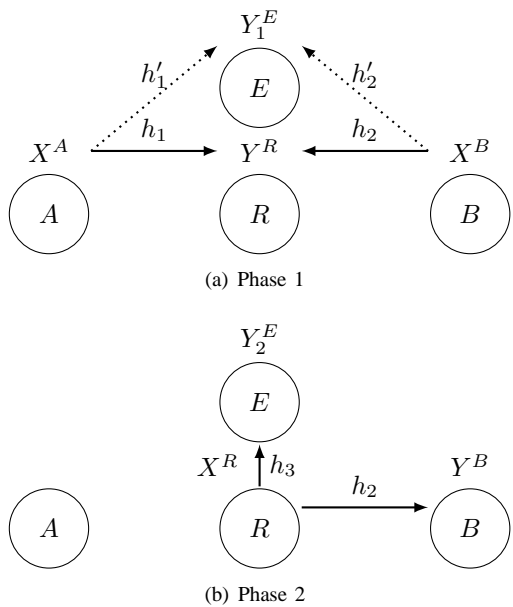

In the second phase, the relay transmits to node $B$, which is also overheard by the eavesdropper.
\begin{eqnarray}\label{eq:ybx}
Y^B &=&h_2X^R+Z^B, \\
Y^E_2&=&h_3X^R+Z_2^E,
\end{eqnarray}
where channel coefficient $h_3 \in \mathbb{R}^+$ and $Z^B, Z_2^E$ are also $N$-dimensional independent Gaussian noise vectors. The model is illustrated in Fig.~\ref{fig:exte}.

The power constraints for the transmission of node $A,B$, and $R$ are given in (\ref{eq:power}).
W.l.o.g., we let all noise vectors have unit variance in each dimension. We assume the power constraints as well as all channel coefficients are revealed to all nodes. In this model, the reliability constraint is still (\ref{eq:reli}). However, the secrecy constraint becomes
\begin{equation}\label{eq:sec2}
\lim_{N \to \infty}\frac{1}{N}I(W^A;Y^E_1, Y^E_2)\leq \delta
\end{equation}
for any chosen $\delta>0$.

\remark{This problem is essentially different from the normal wire-tap type of problems or the problem we introduced in Subsection~\ref{s:model}. The main difference is that the information is leaked to the eavesdropper twice from the source and the relay, respectively. Most of the existing secure transmission problems only focus on preventing the eavesdropper from getting information from one source.}

\subsection{Coding Scheme}

Here, we show that the RB scheme can be straightforwardly applied to achieve a positive secrecy rate if the transmit rate of the $({\bf a}, {\boldsymbol\beta})$ SCF code is set appropriately. The key for the deployment of the RB scheme is that the transmit rate should be chosen such that the eavesdropper can decode the same linear combination that the relay decodes. Then, the reception of the eavesdropper in the second phase will be a degraded version of the first phase, i.e., $H(Y^E_2|Y^E_1)=0$.


Before setting the transmit rate, we first give the definition of the computation rate in this model. Note that the expression of the computation rate is different from (\ref{eq:comr2}), which is due to the non-unit channel coefficients. The code and the transmit process are essentially the same. We define the computation rate in this model for $i \in \{A, B\}$ as
\begin{equation}\label{eq:comr3}
\tilde{R}_{\rm CF}^i({\bf a}, {\boldsymbol\beta},{\bf h}^*)=\frac{1}{2}\log_2(\frac{\beta_i^2 h_i^2 P_i}{{\cal N}({\bf a}, {\boldsymbol\beta},{\bf h}^*)}),
\end{equation}
where 
\begin{equation}
{\cal N}({\bf a}, {\boldsymbol\beta},{\bf h}^*)=\frac{h_A^2h_B^2P_AP_B(a_1\beta_A-a_2\beta_B)^2+(a_1h_A\beta_A)^2P_A+(a_2h_B\beta_B)^2P_B}{h_A^2P_A+h_B^2P_B+1}.
\end{equation}
 Here, ${\bf h}^*=(h_A, h_B) \in \mathbb{R}^+$. Note that ${\cal N}({\bf a}, {\boldsymbol\beta},{\bf h}^*)$ is simply ${\cal N}({\bf a}, {\boldsymbol\beta})$ in (\ref{eq:nabeta}) with all $P_i$ substituted by $h_i^2P_i$.

Now we set the value for the transmit rate $R_t^i({\bf a}, {\boldsymbol\beta}), i\in\{A, B\}$. Firstly, to guarantee $H(Y^E_2|Y^E_1)=0$, we let the relay decode a linear combination $a_1X^A+a_2X^B$ which the eavesdropper can also decode. In this case, the transmission of the relay at Phase 2 will not leak more information since we have a Markov chain $Y_E^1 \to a_1X^A+a_2X^B \to X^R \to Y^E_2$. By the data processing inequality we have $H(Y^E_2|Y^E_1)=0$ and
\begin{equation}\label{eq:sec3}
I(W^A;Y^E_1, Y^E_2) = I(W^A;Y^E_1).
\end{equation}
By \cite{jingge}, there exists a sequence of lattice codes with which the relay and the eavesdropper are both able to decode $a_1X^A+a_2X^B$ if the transmit rate of node $i \in\{A, B\}$ satisfies
\begin{equation}\label{eq:rtee}
R_t^i({\bf a}, {\boldsymbol\beta}) < \min (\tilde{R}_{\rm CF}^i({\bf a}, {\boldsymbol\beta},{\bf h}),\tilde{R}_{\rm CF}^i({\bf a}, {\boldsymbol\beta},{\bf h}')).
\end{equation}
Similarly to (\ref{eq:il4}), the information leakage rate of this code can be bounded by
\begin{equation}\label{eq:ro2}
R_o({\bf a}, {\boldsymbol\beta}) < C({h_1'}^2P_A+{h_2'}^2P_B)-R_t^B({\bf a}, {\boldsymbol\beta}).
\end{equation}

Now, using the RB scheme with the transmit rate of the $({\bf a}, {\boldsymbol\beta})$ SCF code set accordingly to (\ref{eq:rtee}) and the random binning code generated w.r.t. (\ref{eq:ro2}), the reliable and secure transmission is guaranteed. The proof for the reliability and security are identical to the proof we given in Subsection~\ref{ss:rbbs}. We thus have the following theorem for the achievable secrecy rate.

\theorem{In a two-hop channel with an eavesdropper, any secrecy rate satisfying
\begin{eqnarray}
R_s &<&\max_{{\bf a}, {\boldsymbol\beta}} \sum_{i \in \{A,B\}} \left(\min (\tilde{R}_{\rm CF}^i({\bf a}, {\boldsymbol\beta},{\bf h}),\tilde{R}_{\rm CF}^i({\bf a}, {\boldsymbol\beta},{\bf h}'))\right) -C({h_1'}^2P_A+{h_2'}^2P_B)
\end{eqnarray}
is achievable if $P_R \geq \frac{h_1^2}{h_2^2}P_A$ and any secrecy rate satisfying
\begin{eqnarray}
R_s&<&\max_{{\bf a}, {\boldsymbol\beta}}\left\{ \left[\sum_{i \in \{A,B\}} \left(\min (\tilde{R}_{\rm CF}^i({\bf a}, {\boldsymbol\beta},{\bf h}),\tilde{R}_{\rm CF}^i({\bf a}, {\boldsymbol\beta},{\bf h}'))\right)-C({h_1'}^2P_A+{h_2'}^2P_B)\right] \right. \nonumber\\ 
&& \hspace{7cm} \left.\vphantom{\sum_{i \in \{A,B\}}}\frac{C(h_2^2P_R)}{\min (\tilde{R}_{\rm CF}^A({\bf a}, {\boldsymbol\beta},{\bf h}),\tilde{R}_{\rm CF}^A({\bf a}, {\boldsymbol\beta},{\bf h}'))} \right\} \nonumber \\
\end{eqnarray}
is achievable if $P_R <\frac{h_1^2}{h_2^2}P_A$.
}\label{th:ee}

\remark{When $h_1' = h_1$, $h_2' = h_2$, and $P_R \geq \frac{h_1^2}{h_2^2}P_A$, Theorem~\ref{th:ee} mimics our result in Corollary~\ref{co:1}. However, when the eavesdropper has a bad channel, e.g., $h_1' \to 0$ and $h_2' \to 0$, the achievable rate in Theorem~\ref{th:ee} tends to zero, which reflect the sub-optimality of this scheme.}

\begin{figure}[!ht]
\centering
\includegraphics[width=0.5\textwidth]{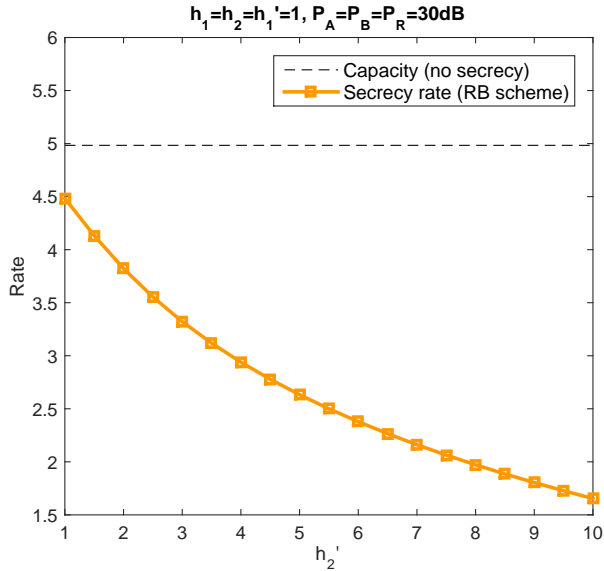}
\caption{Achievable secrecy rate in a two-hop channel with an eavesdropper when $h_1=h_2=h_1'=1$ as a function of $h_2'$.}
\label{fig:exe}
\end{figure}

In Fig.~\ref{fig:exe} we show an example in which the RB scheme achieves a positive rate on a two-hop channel with an eavesdropper. The achievable secrecy rate is identical to the untrusted relay case when $h_1=h_2=h_1'=h_2'=1$ and decreases if the $h_2'$ increases. With our scheme, a large $h_1'$ or a large $h_2$ will both result in small or even no secrecy rate at all.

\section{Conclusion}\label{s:con}

In this paper, we proposed two novel reliable and secure transmission schemes for the two-hop channel with an untrusted relay. These are the first secure transmission schemes that use the scaled compute-and-forward technique. We have shown that when the cooperative jammer and the destination are collocated, both of our schemes achieve relatively good secrecy rates in the high SNR region. Especially, for $P_B= \alpha P_A, P_R= \gamma P_A, \alpha, \gamma \in \mathbb{R}^+$, and $P_A \to \infty$, our schemes are the first upper bound achieving schemes for any $\alpha$ and $\gamma$. Moreover, the LC scheme is the first upper bound achieving scheme if $P_R$ is limited and $P_A$ is not unbounded. In summary, our schemes significantly improve the achievable secrecy rate lower bound and achieve the upper bound in two cases:
1, $P_R$ is limited and $P_A$ does not go to infinity.
2, $P_A, P_B, P_R$ are linearly related and go to infinity.

Also, we proposed another two-hop channel model in which the relay is trusted but there exists an external eavesdropper. We showed that our RB scheme can also be exploited in this model and achieves a positive secrecy rate.



%


\appendices

\section{Proof of Lemma~\ref{lm:rb}}\label{ap:1}
\beginofproof
By analyzing the information leakage rate, we will have
\begin{eqnarray}
\lefteqn{\frac{1}{lN}I({\cal W}^A;{\cal Y}^R)} \nonumber \\
 &=&\frac{1}{lN}(H({\cal W}^A)-H({\cal X}^A,{\cal X}^B|{\cal Y}^R) -H({\cal W}^A|{\cal Y}^R,{\cal X}^A,{\cal X}^B)+H({\cal X}^A,{\cal X}^B|{\cal Y}^R,{\cal W}^A)) \label{eq:il11}\\
&=& \frac{1}{lN}(H({\cal W}^A)-H({\cal X}^A, {\cal X}^B) +I({\cal X}^A,{\cal X}^B;{\cal Y}^R)+H({\cal X}^A,{\cal X}^B|{\cal Y}^R,{\cal W}^A)), \label{eq:il12}
\end{eqnarray} 
where ${\cal W}^A$ is length-$l$ sequence of source messages and ${\cal X}^A$, ${\cal X}^B$, and ${\cal Y}^R$ are length-$\lceil l' \rceil$ sequences of the transmissions of the source, the transmissions of the jammer, and the receptions at the relay, respectively.
The third term in (\ref{eq:il11}) is 0 since the mapping error from the codewords to the source messages is almost zero as we stated in Segment~\ref{sss:re}. Then we focus on the last term $H({\cal X}^A,{\cal X}^B|{\cal Y}^R,{\cal W}^A)$. This term can be upper bounded by Fano's inequality since the relay can determine the transmitted codeword almost surely. The reason is that the size of the bin is chosen accordingly to the information leakage rate.

More precisely, for any transmitted message, by the leaked information, the relay is able to list almost $2^{\lfloor lH(W^A) \rfloor}$ possible ${\cal X}^A$ as candidates from a total of $2^{\lfloor lH(W^A) \rfloor+\lfloor l' N R_o({\bf a}, {\boldsymbol\beta})\rfloor}$ codewords. Then, since the random binning process is independent and uniform, the relay can determine the transmitted codeword almost surely if it also knows the label of the bin when $N,l \to \infty$.

Then, by Fano's inequality, we have
\begin{eqnarray}
\lefteqn{H({\cal X}^A,{\cal X}^B|{\cal Y}^R,{\cal W}^A)}\nonumber \\
&\leq &\frac{1}{lN}+\frac{1}{lN}P_e\log_2 2^{\lfloor lH(W^A) \rfloor+\lfloor l' N R_o({\bf a}, {\boldsymbol\beta})\rfloor} \nonumber \\
&\leq& \epsilon',
\end{eqnarray}
where $P_e,\epsilon' \to 0$ when $N \to \infty$.

Hence, we have
\begin{eqnarray}
\lefteqn{\frac{1}{lN}I({\cal W}^A;{\cal Y}^R)} \nonumber \\
&=& \frac{1}{lN}(H({\cal W}^A)-H({\cal X}^A)-H( {\cal X}^B) +I({\cal X}^A,{\cal X}^B;{\cal Y}^R)+H({\cal X}^A,{\cal X}^B|{\cal Y}^R,{\cal W}^A)) \label{eq:il12} \nonumber \\
&=& \frac{1}{lN}(lH(W^A)-\lfloor lH(W^A)\rfloor -\lfloor l'NR_o({\bf a}, {\boldsymbol\beta})\rfloor -\lceil l' \rceil NR_t^B({\bf a}, {\boldsymbol\beta})+\lceil l' \rceil I(X^A,X^B;Y^R)+\epsilon' \nonumber \\
&<& \frac{1}{lN}(lH(W^A)-\lfloor lH(W^A)\rfloor -\lfloor l'NR_o({\bf a}, {\boldsymbol\beta})\rfloor -\lceil l' \rceil NR_t^B({\bf a}, {\boldsymbol\beta})+\lceil l' \rceil NC(P_A+P_B))+\epsilon' \nonumber \\
&\leq& \frac{1}{lN}(1-\lfloor l' NC(P_A+ P_B)-l'NR_t^B({\bf a}, {\boldsymbol\beta})\rfloor  -\lceil l' \rceil  NR_t^B({\bf a}, {\boldsymbol\beta})+\lceil l' \rceil NC(P_A+P_B))+\epsilon' \nonumber \\
&\leq& \frac{1}{lN}(2-l' N(C(P_A+ P_B)-R_t^B({\bf a}, {\boldsymbol\beta}))  +\lceil l' \rceil  N(C(P_A+P_B)-R_t^B({\bf a}, {\boldsymbol\beta}) ))+\epsilon' \nonumber \\
&\leq& \frac{2}{lN}+\frac{-R_t^B({\bf a}, {\boldsymbol\beta})+C(P_A+P_B)}{l}+\epsilon',
\end{eqnarray} 
which can be made arbitrarily small by choosing sufficiently large $l,N$. Here, the second equality follows from our codebook construction, where $H({\cal X}^A) = \lfloor lH(W^A)\rfloor +\lfloor l'NR_o({\bf a}, {\boldsymbol\beta})\rfloor$ and $H( {\cal X}^B) = \lceil l' \rceil NR_t^B({\bf a}, {\boldsymbol\beta})$. The first inequality follows from the capacity of the Gaussian MAC. The second inequality follows from (\ref{eq:il4}).


\endofproof

\section{Proof of Lemma~\ref{lm:2}}\label{ap:2}
\beginofproof
Firstly, since $X^B$ is independent of $X^A$ and the dithers are known, by  (\ref{eq:rta}) we have 
\begin{eqnarray}
\lefteqn{\frac{1}{N}(H(W^A)-H(X^A)-H(X^B))} \nonumber \\
&=&R_s^A({\bf a}, {\boldsymbol\beta})-R_t^A({\bf a}, {\boldsymbol\beta})-R_t^B({\bf a}, {\boldsymbol\beta})\nonumber \\
&=&-R_e^A({\bf a}, {\boldsymbol\beta})-R_t^B({\bf a}, {\boldsymbol\beta}). \label{eq:il21}
\end{eqnarray}
We can then upper bound the information leakage rate at the relay by
\begin{eqnarray}
\lefteqn{\frac{1}{N} I(W^A; Y^R)}  \nonumber \\
&=& \frac{1}{N}(H(W^A)-H(X^A,X^B|Y^R) -H(W^A|Y^R,X^A,X^B)+H(X^A,X^B|Y^R,W^A)) \nonumber\\
&=& \frac{1}{N}(H(W^A)-H(X^A)-H(X^B) +I(X^A,X^B;Y^R)) +\frac{1}{N}H(X^A,X^B|Y^R,W^A). \nonumber \\
&<& -R_e^A({\bf a}, {\boldsymbol\beta})-R_t^B({\bf a}, {\boldsymbol\beta})+C(P_A+P_B) +\frac{1}{N}H(X^A,X^B|Y^R,W^A).  \label{eq:apinle}
\end{eqnarray}
Here, the second equality follows a similar argument as (\ref{eq:il12}). The inequality follows from (\ref{eq:il21}) and the Gaussian MAC capacity.

By \cite[Theorem 2, 3]{jingge2}, the decoder can reliably decode $X^A$ and $X^B$ from $Y^R$ and $W^A$ with all the dithers, lattices and coefficients by a regular lattice decoding scheme if (\ref{eq:conrsa}) holds. Here, we briefly explain the decoding process.

Firstly, we let the relay decode $a_1(V^A+T^A)+a_2V^B$. By \cite{jingge}, the decoding is successful if (\ref{eq:racf}) holds, which is guaranteed by (\ref{eq:rtcf}). Then, since the relay already knows $W^A$ and the codebook, it knows $T^A$ as well. We let it subtract $a_1T^A$ and decode $V^A$ by treating $a_1(V^A+T^A)+a_2V^B$ as noise. It is proved that the decoding is reliable if
\begin{equation}\label{eq:conrea}
R_e^A({\bf a}, {\boldsymbol\beta})<C(P_A+P_B)-\hat{R}_t^B({\bf a}, {\boldsymbol\beta}),
\end{equation}
which is guaranteed by (\ref{eq:il4}), (\ref{eq:macab}), and (\ref{eq:conrsa}). Then, it can also decode $V^B$ by subtracting $a_1(V^A+T^A)$ from $a_1(V^A+T^A)+a_2V^B$. Hence, both $V^A$ and $V^B$ are decoded and then $X^A$ and $X^B$ are reliably decoded as well.

As a result, by Fano's inequality we have
\begin{eqnarray}
\lefteqn{\frac{1}{N}H(X^A,X^B|Y^R,W^A)} \nonumber \\
&\leq& \frac{1}{N}+\frac{1}{N}P_e\log_2 2^{N(R_e^A({\bf a}, {\boldsymbol\beta})+R_t^B({\bf a}, {\boldsymbol\beta}))} \nonumber \\
&=& \frac{1}{N} + P_e(R_e^A({\bf a}, {\boldsymbol\beta})+R_t^B({\bf a}, {\boldsymbol\beta})),\label{eq:part2}
\end{eqnarray}
where $P_e$ is the probability of decoding errors which tends to 0 when $N \to \infty$. Then, bringing back the expression of information leakage rate in (\ref{eq:apinle}), combining (\ref{eq:il4}), (\ref{eq:macab}), and (\ref{eq:part2}), the information leakage rate can be made arbitrarily small by choosing sufficiently large $N$ and sufficiently small $\delta''$. Hence, we finish the proof.
\endofproof

\section*{Acknowledgment}

This work was supported by ERC Starting Grant 259530-ComCom and by NWO Grant 612.001.107.

The authors would also like to thank Jingge Zhu from Ecole Polytechnique F\'ed\'erale de Lausanne for the valuable discussion.

\end{document}